\documentclass[journal,twoside,web]{ieeecolor}
\usepackage{tmi}
\usepackage{cite}
\usepackage{amsmath,amssymb,amsfonts}
\usepackage{algorithmic}
\usepackage{graphicx}
\usepackage{textcomp}

\usepackage{lipsum}
\usepackage{mathrsfs}
\usepackage{threeparttable}
\usepackage[bookmarks=true]{hyperref}
\usepackage{url}
\usepackage{multirow}
\usepackage{booktabs}
\usepackage{makecell}
\usepackage{colortbl}

\definecolor{gray}{RGB}{128,128,128}
\definecolor{cgreen}{RGB}{197,224,179}
\definecolor{cblue}{RGB}{189,214,238}
\SetLipsumParListSurrounders{\begingroup\color{gray}}{\endgroup}

\newcommand{\newcontent}[1]
{{#1}}

\def\BibTeX{{\rm B\kern-.05em{\sc i\kern-.025em b}\kern-.08em
    T\kern-.1667em\lower.7ex\hbox{E}\kern-.125emX}}
\begin{document}
\title{Domain and Content Adaptive Convolution based Multi-Source Domain Generalization for Medical Image Segmentation}
\author{Shishuai Hu, Zehui Liao, Jianpeng Zhang, and Yong Xia, \IEEEmembership{Member, IEEE}
\thanks{This work was supported in part by the National Natural Science Foundation of China under Grants 62171377 and 61771397. (S. Hu and Z. Liao contributed equally to this work.) (Corresponding author: Yong Xia)}
\thanks{The authors are with the National Engineering Laboratory for Integrated Aero-Space-Ground-Ocean Big Data Application Technology, School of Computer Science and Engineering, Northwestern Polytechnical University, Xi’an 710072, China (e-mail: sshu@mail.nwpu.edu.cn; merrical@mail.nwpu.edu.cn; james.zhang@mail.nwpu.edu.cn; yxia@nwpu.edu.cn).
}
}

\maketitle

\begin{abstract}
The domain gap caused mainly by variable medical image quality renders a major obstacle on the path between training a segmentation model in the lab and applying the trained model to unseen clinical data. To address this issue, domain generalization methods have been proposed, which however usually use static convolutions and are less flexible.
In this paper, we propose a multi-source domain generalization model based on the domain and content adaptive convolution (DCAC) for the segmentation of medical images across different modalities.
Specifically, we design the domain adaptive convolution (DAC) module and content adaptive convolution (CAC) module and incorporate both into an encoder-decoder backbone.
In the DAC module, a dynamic convolutional head is conditioned on the predicted domain code of the input to make our model adapt to the unseen target domain.
In the CAC module, a dynamic convolutional head is conditioned on the global image features to make our model adapt to the test image.
We evaluated the DCAC model against the baseline and four state-of-the-art domain generalization methods on the prostate segmentation, COVID-19 lesion segmentation, and optic cup/optic disc segmentation tasks.
Our results not only indicate that the proposed DCAC model outperforms all competing methods on each segmentation task but also demonstrate the effectiveness of the DAC and CAC modules.
Code is available at \url{https://git.io/DCAC}.

\end{abstract}

\begin{IEEEkeywords}
Domain generalization, medical image segmentation, dynamic convolution, deep learning.
\end{IEEEkeywords}

\section{Introduction}

\begin{figure}[]
  \centering
  \includegraphics[scale=0.37]{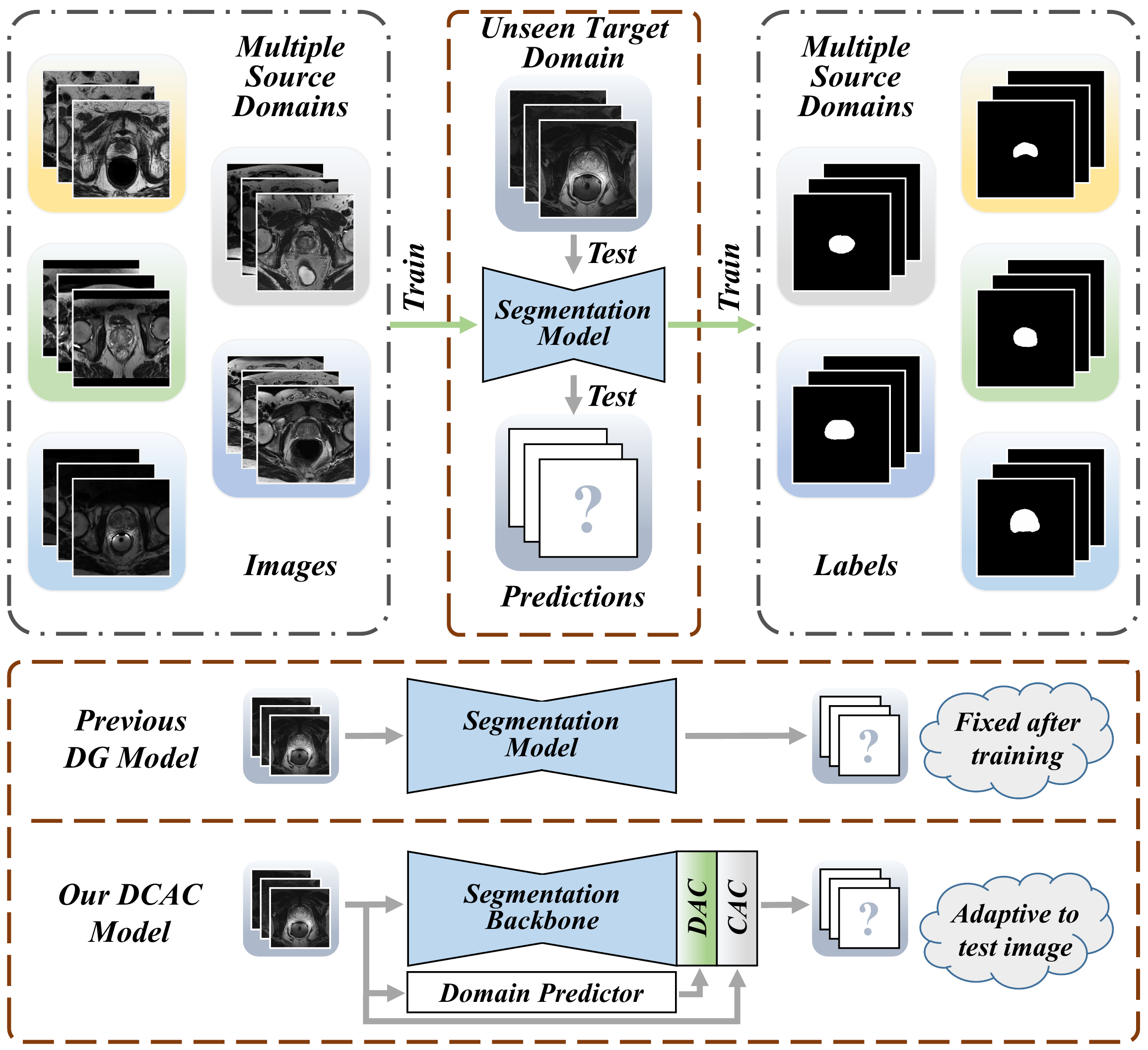}
  \caption{Illustration of (top) multi-source domain generalization, (middle) previous domain generalization model, and (bottom) proposed DCAC model. The images and corresponding segmentation masks from five source domains are highlighted with different background colors. The \textcolor[RGB]{84,130,53}{\textbf{green}} arrows indicate the training process, while the \textcolor[RGB]{127,127,127}{\textbf{gray}} arrows highlight the inference process.
  The segmentation model trained with the data from five source domains is expected to generalize well on the unseen target domain.
  Previously, a domain generalization model is frozen after training and thus uses the same set of parameters to handle various target domain data. In contrast, our DCAC model can adapt to different test images due to the use of dynamic convolutions. DG: Domain generalization.}
  \label{fig:problem}
\end{figure}

\IEEEPARstart{M}{edical} image segmentation is one of the most critical yet challenging steps in computer-aided diagnosis.
Since manual segmentation requires considerable expertise and is time-consuming, expensive, and prone to operator-related bias, automated segmentation approaches are in extremely high demand and have been extensively studied~\cite{LITJENS201760,xie_survey_2021}.

Recent years have witnessed the success of deep learning in medical image segmentation~\cite{falk_u-net_2019,zhou_unet_2020,isensee_nnu-net_2020}.
As a data-driven technique, deep learning requires a myriad amount of annotated training data to alleviate the risk of over-fitting.
However, there is usually a small dataset for medical image segmentation tasks, and this relates to the work required in acquiring the images and then, more importantly, in image annotation~\cite{guo2021learning,wang2021deep,huang2020noise,yang2021towards}.
Due to the small data issue, the i.i.d. assumption, \textit{i.e.}, each training or test data should be drawn independently from an identical distribution, is less likely to be held.
Indeed, the problem of distribution discrepancy between training and test data is particularly severe on medical image segmentation tasks, since the quality of medical images varies greatly over many factors, including different scanners, imaging protocols, and operators~\cite{wang_dofe_2020,liu_feddg_2021}.
As a result, a segmentation model learned on a set of training images may over-fit the data, and hence has a poor generalization ability on test images, which are collected in another medical center and follow a different distribution.
Such undesired performance drop renders a major obstacle on path between the design and clinical application of medical image segmentation tools.

To address this issue, tremendous research endeavors have recently focused on unsupervised domain adaptation (UDA), test time adaptation (TTA), and domain generalization. 
UDA attempts to alleviate the decrease of generalization ability caused by the distribution shift between the labeled source domain (training) data and unlabelled target domain (test) data in three ways.
At the data level, the image-to-image translation is performed to make the quality of source domain data match the quality of target domain data, leading to reduced distribution discrepancy~\cite{yang_fda_2020,chen2020unsupervised,liu2020pdam}.
At the feature level, domain adaptation is achieved by using either adversarial training or feature normalization to extract domain-irrelevant features~\cite{liu_ms-net_2020,shen_domain-invariant_2020}.
At the decision level, various constraints are posed to enforce the consistency between the source domain output and target domain output~\cite{wang2019boundary}.
Despite their promising performance, UDA methods have a limited clinical value due to the requirement of accessing target domain data~\cite{tsai_rsna_2021,roth2021rapid}.

To overcome the limitation of UDA, TTA methods have been proposed to train the segmentation model with the source domain data only, while fine-tuning the trained model with the target domain data at the test time.
It can be accomplished by adding an additional adaptor network to transform~\cite{he2021autoencoder} or normalize~\cite{karani_test-time_2021} the test data and its features to minimize the domain shift at the test time.
Although TTA methods avoid accessing target domain data, they require an extra network to adapt the model to the target data, which increases the spatial and computational complexity.

Domain generalization methods target at boosting the generalization ability of DCNN models and improving their performance in the unseen target domain.
An intuitive solution is to extract domain-invariant features via posing domain-invariant constraints to the model or using adversarial training~\cite{wang_dofe_2020,fan_adversarially_2021,zhou_duplex_2021}. 
Nevertheless, it is not easy to differentiate domain-invariant features from domain-specific ones, especially when the target data distribution is completely unknown.
To increase the diversity of training data, multiple source domains have been increasingly used to replace the single source domain (see Fig.~\ref{fig:problem}).
Multi-source domain generalization methods~\cite{liu_shape-aware_2020,du_metanorm_2021,liu_semi-supervised_2021,liu_feddg_2021} usually employ meta-learning to minimize the generalization gap between the simulated source domain and target domain.
However, if the simulated domain could not cover the unseen target domain, meta-learning-based methods may not perform well.
Alternatively, augmentation-based domain generalization methods~\cite{zhang_generalizing_2020,li_domain_2020} attempt to simulate the target data distribution via augmenting either the source domain data or the features of source data.
Despite their advantages, domain generalization methods still suffer from limited performance, which is attributed mainly to their static nature. Specifically, a domain generalization model is frozen after training and therefore uses the same set of parameters to handle various unseen target data, which have diverse distributions.

In this paper, we propose a multi-source domain generalization model based on the domain and content adaptive convolution (DCAC) for the segmentation of medical images across different modalities.
We adopt an encoder-decoder structure as the backbone and design the domain adaptive convolution (DAC) module and content adaptive convolution (CAC) module.
To adapt our model to the unseen target domain, the DAC module provides a domain-adaptive head, whose parameters are dynamically generated by the domain-aware controller based on the estimated domain code of the input.
To adapt our model to each test image, the CAC module has a content-adaptive head, whose parameters are dynamically produced by the content-aware controller based on the global image features.
We have evaluated the proposed DCAC model on three medical image segmentation benchmarks, including the prostate segmentation in MRI scans from six domains, COVID-19 lung lesion segmentation in CT scans from four domains, and optic cup (OC)/optic disc (OD) segmentation in fundus images from four domains.

Our contributions are three-fold.
\begin{itemize}
\item We used the domain-discriminative information embedded in the encoder feature maps to generate the domain code of each input image, which establishes the relationship between multiple source domains and the unseen target domain.
\item We designed the dynamic convolution-based DAC module and CAC module, which respectively enable our DCAC model to adapt not only to the unseen target domain but also to each test image.
\item We presented extensive experimental results, which demonstrate not only the effectiveness of DAC and CAC modules but also the superiority of our DCAC model over state-of-the-art domain generalization techniques on three medical image segmentation tasks.
\end{itemize}

\section{Related Work}
\subsection{Domain Generalization for Medical Image Segmentation}
Domain generalization methods designed for medical image segmentation can be roughly categorized into augmentation-based, meta-learning-based, and domain-invariant feature learning approaches.
\textbf{Augmentation-based methods}, such as the deep stacked transformation~\cite{zhang_generalizing_2020}, simulate the distribution of target domain data by augmenting the source domain data. The linear-dependency domain generalization methods~\cite{li_domain_2020,gu2021domain} perform the augmentation in the feature space, aiming to simulate the distribution of features instead of the distribution of data.
With the recent advance of the episodic training strategy for domain generalization in computer vision~\cite{li_episodic_2019}, many \textbf{meta-learning-based methods} have been developed to generalize medical image segmentation models to unseen domains~\cite{liu_semi-supervised_2021,li_domain_2021}.
For example, a shape-aware meta-learning scheme~\cite{liu_shape-aware_2020}, which takes the incomplete shape and ambiguous boundary of prediction masks into consideration, was proposed to improve the model generalization for prostate MRI segmentation.
In another example, the continuous frequency space interpolation was combined with the episodic training strategy to achieve further performance gains in cross-domain retinal fundus image segmentation and prostate MRI segmentation~\cite{liu_feddg_2021}.
Although these methods work well on specific tasks using elaborately tuned parameters, their performance degrades substantially on the target domain when there are only few source domains. 
Given this, 
\textbf{domain-invariant feature learning methods}~\cite{onofrey_generalizable_2019} have been proposed.
Zhao \textit{et al.}~\cite{zhao_robust_2021} adopted domain adversarial learning and mix-up to improve white matter hyperintensity prediction on an unseen target domain.
Wang \textit{et al.}~\cite{wang_dofe_2020} built a domain knowledge pool to store domain-specific prior knowledge and then utilized domain attribute to aggregate features from different domains.

Different from these methods, the proposed DCAC model uses dynamic convolutions whose parameters are generated by a controller according to the features of an input image, and thus is able to adapt to the test image from an unknown domain.

\subsection{Dynamic Convolutions}

\begin{figure}[]
    \centering
    \includegraphics[scale=0.7]{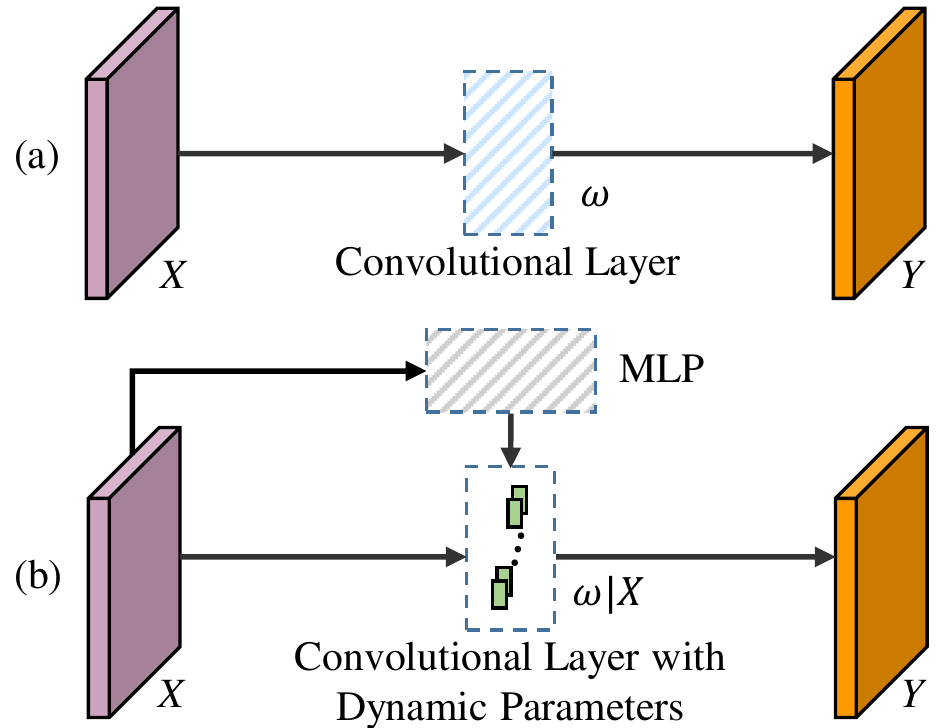}
    \caption{Comparison between traditional convolution and dynamic convolution: (a) the feature map $X$ is processed by a traditional convolutional layer, whose parameters $\omega$ are learned during training; and (b) the feature map $X$ is processed by a dynamic convolutional layer, whose parameters are generated by a multilayer perceptron (MLP) and conditioned on $X$, \textit{i.e.}, $\omega|X$.
    }
    \label{fig:dynamic_convolution}
\end{figure}

The traditional convolution suffers from limited flexibility, since its parameters $\omega$ are learned during training and fixed during inference, regardless of the variations of input, task, and domain (see Fig.~\ref{fig:dynamic_convolution} (a)).
To address this issue, the dynamic convolution has been proposed. Specifically, another network (\textit{e.g.}, an MLP) is employed to generate the convolutional parameters $\omega$ based on various conditions (\textit{e.g.}, the input $X$), and the convolutions with dynamically generated parameters (e.g., $\omega|X$) are then used to process the input (see Fig.~\ref{fig:dynamic_convolution} (b)). 
Since the parameters $\omega$ can change with respect to the current input, task, and/or image domain during inference, the dynamic convolution is far more flexible than its traditional counterpart.
Therefore, various dynamic convolutions have been increasingly studied and used in the field of computer vision~\cite{he_dynamic_2019,pang_hierarchical_2020,zhou_decoupled_2021,han_dynamic_2021}.
A dynamic convolutional layer, in which the filters are generated conditioned on the input image, was proposed for short-range weather prediction based on radar images~\cite{klein_dynamic_2015}.
The dynamic convolutions, whose parameters are generated conditioned on each target instance, were also integrated to the mask head of an instance segmentation network to improve the accuracy and inference speed~\cite{tian_conditional_2020}.
In our previous work, we proposed a convolutional neural network with a dynamic segmentation head, which can be trained on partially labelled abdominal CT scans and be applied to the adaptive segmentation of multiple organs and tumors~\cite{zhang_dodnet_2021}. In the dynamic head, convolutional parameters are generated conditioned on the combination of a task encode and global image features. 
By contrast, the DCAC model proposed in this study aims to filter out domain-specific features dynamically and be aware of the content of an input image. Therefore, DCAC contains a DAC head and a CAC head. The DAC head is composed of only one dynamic convolutional layer, in which the dynamic filters are conditioned on the domain code; while the CAC head contains three dynamic convolutional layers, in which the dynamic filters are conditioned on the global features of an input image.

\section{Method}
\subsection{Problem Definition and Method Overview}
Let a set of $K$ source domains be denoted by $D_s=\{(x_{ki}, y_{ki})_{i=1}^{N_k}\}_{k=1}^K$, where $x_{ki}$ is the $i$-th image in the $k$-th source domain, and $y_{ki}$ is the segmentation mask of $x_{ki}$.
Our goal is to train a segmentation model $F_\theta: x \to y$ on $D_s$, which can generalize well to an unseen target domain $D_t=(x_{i})_{i=1}^{N_t}$.

\begin{figure*}[]
    \centering
    \includegraphics[scale=0.48]{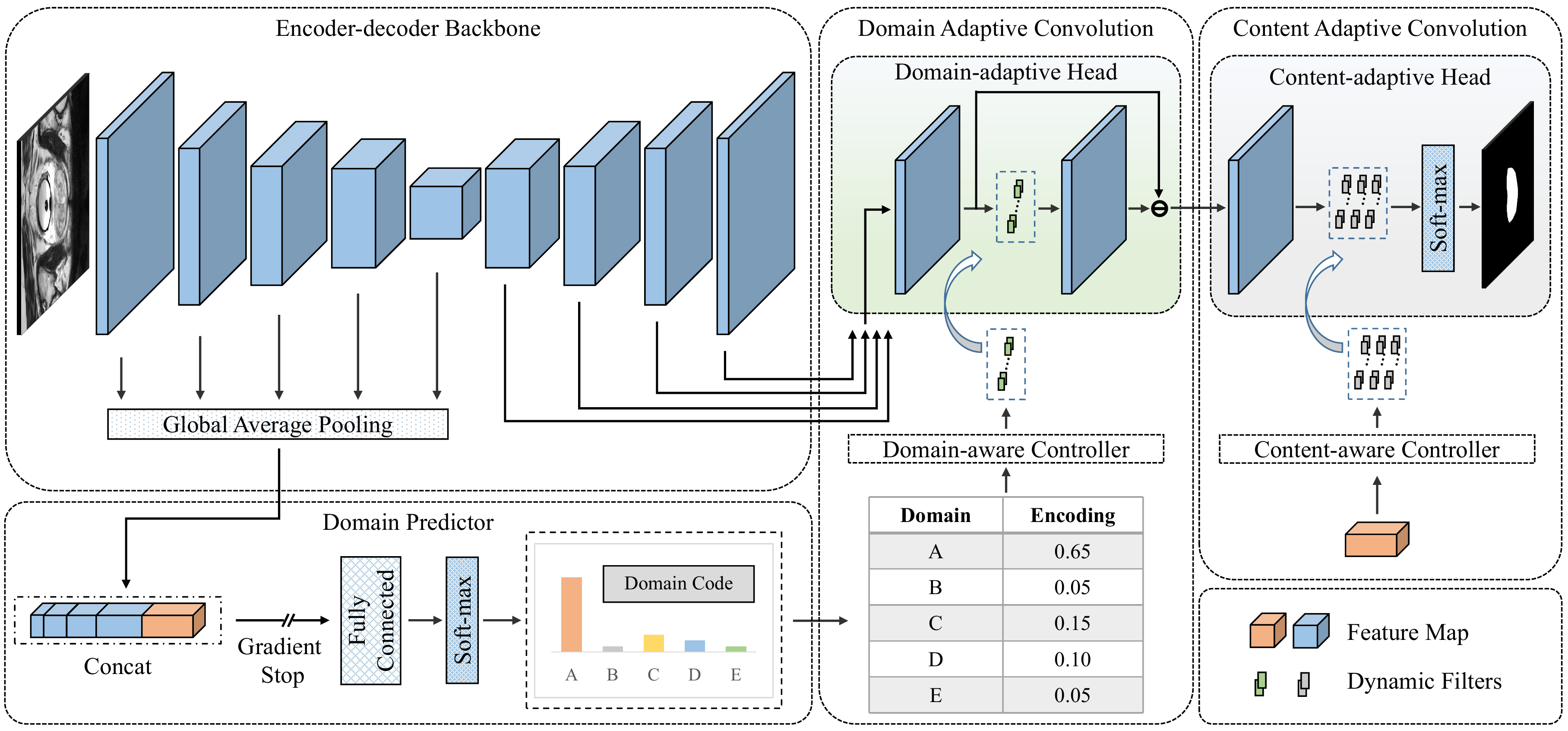}
    \caption{Architecture of the proposed DCAC model. The feature map in orange color represents $GAP(f_E^N)$, \textit{i.e.}, the output of the $N$-th encoder block after global average pooling. The traditional convolutions are omitted for simplicity. The dashed boxes with dynamic filters represent dynamic convolutions.
    }
    \label{fig:overview}
\end{figure*}

The proposed DCAC model is an encoder-decoder structure~\cite{falk_u-net_2019} equipped with a domain predictor, a domain-aware controller, a content-aware controller, and a series of domain-adaptive heads and content-adaptive heads.
The workflow of this model consists of four steps.
First, the feature map produced by each encoder layer is aggregated using Global Average Pooling (GAP) and concatenated together to be fed to the domain predictor to generate the domain code. 
Second, based on the generated domain code, the domain-aware controller predicts the parameters of the domain-adaptive head. 
Third, the content-aware controller uses the final output of the encoder as its input to generate the parameters of the content-adaptive head. 
Finally, according to the deep supervision strategy, the output of each decoder layer is fed sequentially to a domain-adaptive head and a content-adaptive head, which predict the segmentation result on a pixel-by-pixel basis.
The diagram of our DCAC model is shown in Fig.~\ref{fig:overview}. We now delve into its details.

\subsection{Encoder-decoder Backbone}
The backbone used in our DCAC model is a U-shape structure that has an encoder and a decoder, each being composed of $N=4\sim6$ blocks depending on the given segmentation task.
Each encoder block contains two convolutional layers with a kernel size of 3, and the first layer has a stride of 2 to downsample the feature map, except for the first encoder block.
Each layer is followed by instance normalization and the LeakyReLU activation.
In the encoder, the number of filters is set to 32 in the first layer, then doubled in each next block, and finally fixed with 320 when it becomes larger than 256~\cite{isensee_nnu-net_2020}.
The computation in each encoder block can be formally expressed as
\begin{equation}
\begin{aligned}
f_E^i = {Enc}^i(f_E^{i-1};\theta_E^i), \quad i=1,2,\cdots, N
\end{aligned}
\end{equation}
where $\theta_E^i$ represents the parameters of the $i$-th encoder block ${Enc}^i$, $f_E^i$ is the feature map produced by ${Enc}^i$, and $f_E^0 = x^i$ is the input image.

Symmetrically, the decoder upsamples the feature map and refines it gradually.
In each decoder block, the transposed convolution with a stride of 2 is used to improve the resolution of input feature maps, and the upsampled feature map is concatenated with the corresponding low-level feature map from the encoder before being further processed by two convolutional layers. 
The computation in each decoder block can be formally expressed as
\begin{equation}
\begin{aligned}
f_D^i = {Dec}^i(C(f_E^i, U(f_D^{i+1}));\theta_D^i),\ i=N-1,N-2\cdots, 1
\end{aligned}
\end{equation}
where $U(\cdot)$ represents upsampling, $C(\cdot)$ represents concatenation, $\theta_D^i$ represents the parameters of the $i$-th decoder block ${Dec}^i$, $f_D^i$ is the feature map produced by ${Dec}^i$, and $f_D^{N}=f_E^N$.

With this encoder-decoder architecture, multiscale encoder feature maps $\{f_E^i\}_{i=1}^N$ and multiscale decoder feature maps $\{f_D^i\}_{i=1}^{N-1}$ can be generated.
It is expected that $\{f_E^i\}_{i=1}^N$ are domain-sensitive and can be utilized to calculate the probabilities of belonging to source domains of the input image.
Meanwhile, $\{f_D^i\}_{i=1}^{N-1}$ are expected to be rich-semantic and not subjected to a specific domain, \textit{i.e.}, containing the semantic information of domains and target tasks.

\subsection{Domain Adaptive Convolution}
Due to the discrepancy between source domains and the unseen target domain, the encoder-decoder backbone trained with source domain images may not be optimal for target domain images. Therefore, we equipped the backbone with domain-adaptive heads, in which the filters are variable and adaptive to the domain of the input image in the inference stage. For each test image, its probabilities of belonging to source domains, known as a domain code, are calculated by the domain predictor and fed to the domain-aware controller to generate the filters used in the domain-adaptive heads (see Fig.~\ref{fig:overview}).

\subsubsection{Domain Predictor}
Although the target domain is not identical to each source domain, an image in the target domain may similar to those in one or more source domains. And such `domain attribute' of the image can be used as the clue to guide the adaptive processing of it.
Therefore, we design the domain predictor to predict the probability of each target domain image belonging to each source domain.

The domain predictor takes multi-scale encoder feature maps $\{f_E^i\}_{i=1}^N$ as its input.
Each feature map $f_E^i$ is aggregated with GAP, and the aggregated features at all scales are then concatenated into a vector.
To predict the domain code of the input image, the vector is fed to a classification module, which is composed of a fully-connected layer $FC(\cdot)$ and a soft-max layer $SM(\cdot)$.
The calculation of each domain code can be formally expressed as
\begin{equation}
\begin{aligned}
\mathscr{D}^p = SM((FC(C(GAP(f_E^1), \cdots, GAP(f_E^N))); \theta_{FC})),
\end{aligned}
\end{equation}
where $\theta_{FC}$ represents the parameters of $FC(\cdot)$.
The domain code $\mathscr{D}^p$ is a $K$-dimensional vector that satisfies $\sum_{k=1}^{K}\mathscr{D}_k^p = 1$.
During training, since each input image is sampled from one of $K$ source domains, the ground truth domain code that supervises the training of domain predictor is a one-hot $K$-dimensional vector.
Note that image segmentation and domain prediction are different tasks, though using the same set of features extracted by encoder blocks. To avoid the interference with the image segmentation performance caused by domain prediction, we adopt the gradients truncation strategy to stop the gradients back propagated from the fully-connected layer in the domain predictor(see Fig.~\ref{fig:overview}).

\subsubsection{Domain-aware Controller}
We use a single traditional convolutional layer as the domain-aware controller $\phi_d(\cdot)$, which maps the domain code to the parameters $\omega_d$ of the filters in the domain adaptive head.
Such mapping can be formally expressed as
\begin{equation}
\begin{aligned}
\omega_d = \phi_d(\mathscr{D}^p; \theta_\phi^d)
\end{aligned}
\end{equation}
where $\theta_\phi^d$ represents the parameters in this controller.

\subsubsection{Domain-adaptive Head}
A lightweight domain-adaptive head is designed to enable dynamic convolutions, which are responsive to specific domains.
This head contains a traditional convolutional layer and a dynamic convolutional layer, both using filters with a kernel size of 1.
The traditional layer reduces the channels of the input feature map to $K\times C$, where $C$ is the number of segmentation classes. 
Since there exists a skip connection to enforce residual learning, the output of the dynamic layer has $K\times C$ channels, too.
Therefore, there are totally $(K\times C)^2 + (K\times C)$ parameters in the dynamic layer, which are generated dynamically by the domain-aware controller $\phi_d(\cdot)$ conditioned on the domain code $\mathscr{D}^p$.
\newcontent{Thanks to the superiority of dynamic convolutions, the parameters in the dynamic layer are domain-specific, and the output of the dynamic layer can represent the domain-specific feature.}

To accelerate the convergence of our DCAC model, we adopt the multi-scale supervision strategy.
Given the feature map $f_D^i$ generated by the $i$-th decoder block, the output of the domain-adaptive head is computed as
\begin{equation}
\begin{aligned}
f_{DAC}^i = {Conv}_O^i(f_D^i)-{Conv}_O^i(f_D^i)*\omega_d,& \\
i=N-1,N-2\cdots, 1&
\end{aligned}
\end{equation}
where $*$ represents the convolution, and ${Conv}_O^i(\cdot)$ is the traditional convolutional layer.

\subsection{Content Adaptive Convolution}
The proposed DCAC model is expected to adapt not only to the unseen test domain but also to each test image. Therefore, we equipped our segmentation backbone with content adaptive convolutions, which are implemented using a content-adaptive head whose parameters are generated dynamically by a content-aware controller.

\subsubsection{Content-aware Controller}
The content-aware controller is a traditional convolutional layer, denoted by $\phi_c$.
The input of this controller is the global image representation, which is the feature map generated by the encoder (\textit{i.e.}, the output $f_E^N$ of the $N$-th encoder block) and aggregated by global average pooling.
The output is the ensemble of parameters of the content-adaptive head, which can be formally expressed as
\begin{equation}
\begin{aligned}
\omega_c = \phi_c(GAP(f_E^N); \theta_\phi^c)
\end{aligned}
\end{equation}
where $\theta_\phi^c$ represents the parameters of the controller $\phi_c$.

\subsubsection{Content-adaptive Head}
The content-adaptive head, which is placed after the domain-adaptive head, contains three stacked dynamic convolutional layers using filters with a kernel size of 1.
The first two layers have $K\times C$ channels, and the last layer has $C$ channels.
Thus there are totally $2\times ((K\times C)^2 + (K\times C)) + ((K\times C)\times C + C)$ dynamic parameters in this head. These parameters, denoted by $\omega_c=\{\omega_{c1}, \omega_{c2}, \omega_{c3}\}$, are generated by the controller $\phi_c$ according to the globally aggregated image feature map $f_E^N$.

The content-adaptive head uses the output of domain-adaptive head $f_{DAC}^i$ as its input. This head acts as a pixel classifier, performing image segmentation via predicting class labels on a pixel-by-pixel basis.
The computation of segmentation result $p^i$ can be formally expressed as
\begin{equation}
\begin{aligned}
p^i = SM(((f_{DAC}^i*\omega_{c1})*\omega_{c2})*\omega_{c3}),&\\
i=N-1, N-2,\cdots, 1&
\end{aligned}
\end{equation}
where $SM(\cdot)$ represents the soft-max operation.

\subsection{Training and Test}
\subsubsection{Training}
Besides image segmentation, the proposed DCAC model also performs domain classification using the domain predictor. For the classification task, the objective is the cross-entropy loss, which can be calculated as
\begin{equation}
\begin{aligned}
\mathcal{L}_{cls} = - \sum_{k=1}^Kd_k\log(d_k^p)
\end{aligned}
\end{equation}
where $d_k$ is the domain label, and $d_k^p$ is the soft-max probability of belonging to the $k$-th domain.

For the segmentation task, the Dice loss and cross-entropy loss are used jointly as the objective. The segmentation loss at each scale can be calculated as
\begin{equation}
\begin{aligned}
\mathcal{L}_{seg}^i=& 1-\frac{2 \sum_{v=1}^{V} p_{v}^i y_{v}^i}{\sum_{v=1}^{V}\left(p_{v}^i+y_{v}^i+\epsilon\right)} \\
&-\sum_{v=1}^{V}\left(y_{v}^i \log p_{v}^i+\left(1-y_{v}^i\right) \log \left(1-p_{v}^i\right)\right)
\end{aligned}
\end{equation}
where $p_v^i$ and $y_v^i$ denote the prediction and ground truth of the $v$-th voxel in the output of the $i$-th decoder block, $V$ represents the number of voxels, and $\epsilon$ is a smooth factor to avoid dividing by 0.

Since deep supervision is used, the total loss is defined as follows
\begin{equation}
\begin{aligned}
\mathcal{L} = \mathcal{L}_{cls} + \sum_{i=1}^{N-1} \omega^i\mathcal{L}_{seg}^i
\end{aligned}
\end{equation}
where $\omega^i$ is a weighting vector that enables higher resolution output to contribute more to the total loss~\cite{isensee_nnu-net_2020}.

\subsubsection{Test}
During inference, given a test image $x$, the multiscale encoder feature maps $\{f_E^i\}_{i=1}^N$ and multiscale decoder feature maps $\{f_D^i\}_{i=1}^{N-1}$ can be produced by the trained encoder-decoder backbone.
Based on $\{f_E^i\}_{i=1}^N$, the trained domain predictor can generate a $K$-dimensional domain code.
Based on the code, the trained domain-aware controller can generate the parameters for the domain-adaptive head.
Meanwhile, based on the feature map produced by the last encoder block (\textit{i.e.}, $f_E^N$), the content-aware controller can generate the parameters for the content-adaptive head.
Finally, the feature map produced by the decoder is fed sequentially to the domain-adaptive dynamic head and content-adaptive head to generate the segmentation result.
Note that deep supervision is carried out only in the training stage and the segmentation is not performed at coarse scales in the test stage. 

\begin{table*}[]
\centering
\caption{Statistics of three datasets used for this study.}
\label{tab:datasets}
\setlength\tabcolsep{10pt}
\renewcommand{\arraystretch}{1.2} 
\begin{threeparttable}
\begin{tabular}{lcccc}
\hline
\hline
Task & Modality & Number of Domains & Cases in Each Domain & Total Cases \\ \hline
Prostate Segmentation & MRI & 6 & 30; 30; 19; 13; 12; 12 & 116 \\
COVID-19 Segmentation & CT  & 4 & 28; 19; 58; 15 & 120 \\
OC/OD Segmentation & Color Fundus Image & 4 & 50/51; 99/60; 320/80; 320/80\tnote{*} & 789/281\tnote{*}\\ \hline \hline
\end{tabular}
\begin{tablenotes}
\footnotesize
\item[*] Data split (training/test cases) was provided by~\cite{wang_dofe_2020}.
\end{tablenotes}
\end{threeparttable}
\end{table*}

\begin{table*}[]
\centering
\caption{
\newcontent{Performance (mean$\pm$standard deviation) of our DCAC model and five competing models in prostate segmentation. In each column, the best result is shown with \textbf{bold}, and if the result of DCAC and best competing result are statistically different, the cell is highlighted in green; otherwise the cell is in blue. The results of Intra-domain are also displayed for reference.}
}
\label{tab:prostate}
\renewcommand{\arraystretch}{1.2} 
\setlength\tabcolsep{4pt}
\begin{tabular}{l|cccccccc}
\hline \hline
\multirow{2}{*}{Models} & \multicolumn{2}{c}{Domain 1} & \multicolumn{2}{c}{Domain 2} & \multicolumn{2}{c}{Domain 3} & \multicolumn{2}{c}{Domain 4} \\ 
\cline{2-9}
          & DSC$\uparrow$   & ASD$\downarrow$   & DSC$\uparrow$   & ASD$\downarrow$   & DSC$\uparrow$   & ASD$\downarrow$   & DSC$\uparrow$   & ASD$\downarrow$   \\ \hline
Intra-domain & 89.53 & 1.39 & 88.42 & 1.44 & 87.65 & 1.67 & 83.01 & 3.58 \\
\hline
DeepAll & 89.16\newcontent{$\pm$1.01} & 2.09\newcontent{$\pm$0.97} & 87.31\newcontent{$\pm$1.93} & 1.27\newcontent{$\pm$0.70} & 74.12\newcontent{$\pm$3.84} & 3.02\newcontent{$\pm$0.85} & 88.85\newcontent{$\pm$0.97} & 2.36\newcontent{$\pm$0.58} \\
BigAug~\cite{zhang_generalizing_2020} & 90.68\newcontent{$\pm$0.77} & 1.80\newcontent{$\pm$0.76} & 89.52\newcontent{$\pm$0.83} & 1.00\newcontent{$\pm$0.26} & 84.86\newcontent{$\pm$2.51} & \cellcolor{cgreen}1.86\newcontent{$\pm$0.14} & 89.04\newcontent{$\pm$0.71} & 1.59\newcontent{$\pm$0.24}  \\
SAML~\cite{liu_shape-aware_2020} & 91.00\newcontent{$\pm$0.83} & \cellcolor{cgreen}1.26\newcontent{$\pm$0.24} & 89.26\newcontent{$\pm$0.65} & 1.12\newcontent{$\pm$0.30} & \cellcolor{cblue}85.76\newcontent{$\pm$1.67} & 1.87\newcontent{$\pm$0.11} & \cellcolor{cblue}\textbf{89.60\newcontent{$\pm$0.32}} & \cellcolor{cblue}\textbf{1.21\newcontent{$\pm$0.41}} \\
FedDG~\cite{liu_feddg_2021} & \cellcolor{cgreen}91.41\newcontent{$\pm$0.70} & 1.29\newcontent{$\pm$0.16} & \cellcolor{cblue}89.95\newcontent{$\pm$1.53} & \cellcolor{cblue}0.97\newcontent{$\pm$0.21} & 85.10\newcontent{$\pm$1.83} & 2.63\newcontent{$\pm$0.53} & 89.13\newcontent{$\pm$0.87} & 1.51\newcontent{$\pm$0.27}  \\
DoFE~\cite{wang_dofe_2020} & 89.79\newcontent{$\pm$0.66} & 1.33\newcontent{$\pm$0.14} & 87.42\newcontent{$\pm$2.21} & 1.57\newcontent{$\pm$0.51} & 84.90\newcontent{$\pm$2.13} & 2.13\newcontent{$\pm$0.46} & 88.56\newcontent{$\pm$0.94} & 1.52\newcontent{$\pm$0.42} \\
\hline
Ours (DCAC) & \cellcolor{cgreen}\textbf{91.76\newcontent{$\pm$0.48}} & \cellcolor{cgreen}\textbf{0.98\newcontent{$\pm$0.14}} & \cellcolor{cblue}\textbf{90.51\newcontent{$\pm$0.41}} & \cellcolor{cblue}\textbf{0.89\newcontent{$\pm$0.13}} & \cellcolor{cblue}\textbf{86.30\newcontent{$\pm$1.04}} & \cellcolor{cgreen}\textbf{1.77\newcontent{$\pm$0.13}} & \cellcolor{cblue}89.13\newcontent{$\pm$0.82} & \cellcolor{cblue}1.53\newcontent{$\pm$0.49} \\ 
\hline \hline
\end{tabular}

\begin{tabular}{l|cccc|cc}
\hline \hline
\multirow{2}{*}{Models} & \multicolumn{2}{c}{Domain 5} & \multicolumn{2}{c}{Domain 6} & \multicolumn{2}{|c}{Average} \\ 
\cline{2-7}
          & DSC$\uparrow$   & ASD$\downarrow$   & DSC$\uparrow$   & ASD$\downarrow$   & DSC$\uparrow$   & ASD$\downarrow$   \\ \hline
Intra-domain &83.39 & 2.99 & 84.97 & 2.00 & 86.16 & 2.18 \\
\hline
DeepAll & 83.22\newcontent{$\pm$2.53} & 3.51\newcontent{$\pm$0.76} & 88.39\newcontent{$\pm$1.34} & 1.67\newcontent{$\pm$0.49} & 85.18 & 2.32 \\
BigAug~\cite{zhang_generalizing_2020} & 73.24\newcontent{$\pm$5.94} & 5.94\newcontent{$\pm$1.43} & 89.10\newcontent{$\pm$0.89} & 1.16\newcontent{$\pm$0.31} & 86.07 & 2.23  \\
SAML~\cite{liu_shape-aware_2020} & 81.60\newcontent{$\pm$3.76} & 3.29\newcontent{$\pm$0.72} & 89.91\newcontent{$\pm$0.61} & 0.96\newcontent{$\pm$0.23} & 87.86 & 1.62 \\
FedDG~\cite{liu_feddg_2021} & 76.69\newcontent{$\pm$4.49} & 4.52\newcontent{$\pm$0.95} & \cellcolor{cblue}\textbf{90.63\newcontent{$\pm$0.55}} & \cellcolor{cblue}1.03\newcontent{$\pm$0.17} & 87.15 & 1.99  \\
DoFE~\cite{wang_dofe_2020} & \cellcolor{cgreen}\textbf{86.47\newcontent{$\pm$2.29}} & \cellcolor{cblue}\textbf{1.93\newcontent{$\pm$0.64}} & 87.72\newcontent{$\pm$1.71} & 1.33\newcontent{$\pm$0.40} & 87.48 & 1.64 \\
\hline
Ours (DCAC) & \cellcolor{cgreen}83.39\newcontent{$\pm$2.57} & \cellcolor{cblue}2.46\newcontent{$\pm$0.66} & \cellcolor{cblue}90.56\newcontent{$\pm$0.47} & \cellcolor{cblue}\textbf{0.85\newcontent{$\pm$0.19}} & \textbf{88.61} & \textbf{1.41} \\ 
\hline \hline
\end{tabular}
\end{table*}

\begin{table*}[]
\caption{
\newcontent{Performance (mean$\pm$standard deviation) of our DCAC model and five competing models in COVID-19 lesion segmentation. In each column, the best result is shown with \textbf{bold}, and if the result of DCAC and best competing result are statistically different, the cell is highlighted in green; otherwise the cell is in blue. The results of Intra-domain are also displayed for reference.}
}
\label{tab:covid19}
\centering
\setlength\tabcolsep{4pt}
\renewcommand{\arraystretch}{1.2} 
\begin{tabular}{l|cccccccc|cc}
\hline \hline
\multirow{2}{*}{Models} & \multicolumn{2}{c}{Domain 1} & \multicolumn{2}{c}{Domain 2} & \multicolumn{2}{c}{Domain 3} & \multicolumn{2}{c}{Domain 4} & \multicolumn{2}{|c}{Average} \\ 
\cline{2-11}
          & DSC$\uparrow$   & ASD$\downarrow$   & DSC$\uparrow$   & ASD$\downarrow$   & DSC$\uparrow$   & ASD$\downarrow$   & DSC$\uparrow$   & ASD$\downarrow$   & DSC$\uparrow$   & ASD$\downarrow$  \\ \hline
Intra-domain & 62.49 & 21.05 & 51.34 & 21.08 & 70.49 & 6.02 & 62.01 & 8.14 & 61.58 & 14.07  \\
\hline
DeepAll & 63.09\newcontent{$\pm$1.74} & 19.13\newcontent{$\pm$3.71} & 60.87\newcontent{$\pm$2.66} & 19.44\newcontent{$\pm$1.99} & 66.40\newcontent{$\pm$2.64} & 12.21\newcontent{$\pm$2.17} & 62.57\newcontent{$\pm$2.18} & 9.39\newcontent{$\pm$2.02} & 63.23 & 15.04  \\
BigAug~\cite{zhang_generalizing_2020} & 63.55\newcontent{$\pm$1.52} & 18.09\newcontent{$\pm$3.23} & 59.57\newcontent{$\pm$3.11} & 19.53\newcontent{$\pm$2.43} & 67.19\newcontent{$\pm$1.59} & 13.20\newcontent{$\pm$2.36} & 64.39\newcontent{$\pm$1.77} & 9.39\newcontent{$\pm$1.81} & 63.68 & 15.05  \\
SAML~\cite{liu_shape-aware_2020} & 63.98\newcontent{$\pm$1.33} & 15.96\newcontent{$\pm$1.77} & \cellcolor{cgreen}61.39\newcontent{$\pm$1.64} & 18.97\newcontent{$\pm$2.78} & 67.19\newcontent{$\pm$1.72} & 12.87\newcontent{$\pm$2.43} & \cellcolor{cblue}65.38\newcontent{$\pm$1.39} & 9.39\newcontent{$\pm$1.63} & 64.49 & 14.30  \\
FedDG~\cite{liu_feddg_2021} & 63.97\newcontent{$\pm$1.41} & 17.68\newcontent{$\pm$2.80} & 60.88\newcontent{$\pm$2.72} & \cellcolor{cgreen}17.85\newcontent{$\pm$2.29} & 66.96\newcontent{$\pm$1.61} & 13.10\newcontent{$\pm$2.65} & 64.98\newcontent{$\pm$1.32} & \cellcolor{cgreen}9.30\newcontent{$\pm$1.87} & 64.20 & 14.48  \\
DoFE~\cite{wang_dofe_2020} & \cellcolor{cblue}\textbf{64.76\newcontent{$\pm$1.47}} & \cellcolor{cgreen}\textbf{12.43\newcontent{$\pm$2.17}} & 61.11\newcontent{$\pm$1.82} & 18.56\newcontent{$\pm$2.51} & \cellcolor{cblue}67.46\newcontent{$\pm$1.34} & \cellcolor{cgreen}11.74\newcontent{$\pm$1.52} & 65.05\newcontent{$\pm$1.11} & 9.71\newcontent{$\pm$2.14} & 64.60 & 13.11  \\ \hline
Ours (DCAC) & \cellcolor{cblue}64.03\newcontent{$\pm$1.67} & \cellcolor{cgreen}17.05\newcontent{$\pm$1.95} & \cellcolor{cgreen}\textbf{62.52\newcontent{$\pm$1.57}} & \cellcolor{cgreen}\textbf{15.38\newcontent{$\pm$1.76}} & \cellcolor{cblue}\textbf{67.87\newcontent{$\pm$1.25}} & \cellcolor{cgreen}\textbf{10.63\newcontent{$\pm$1.49}} & \cellcolor{cblue}\textbf{65.96\newcontent{$\pm$1.27}} & \cellcolor{cgreen}\textbf{7.98\newcontent{$\pm$1.59}} & \textbf{65.10} & \textbf{12.76}  \\ \hline \hline
\end{tabular}
\end{table*}

\begin{table*}[]
\caption{
\newcontent{Performance (OC, OD) of our DCAC model and five competing models in OC/OD segmentation. In each column, the best result is shown with \textbf{bold}, and if the result of DCAC and best competing result are statistically different, the cell is highlighted in green; otherwise the cell is in blue. The results of Intra-domain are also displayed for reference.}
}
\label{tab:oc/od}
\centering
\setlength\tabcolsep{3pt}
\renewcommand{\arraystretch}{1.2} 
\begin{tabular}{l|cccc}
\hline \hline
\multirow{2}{*}{Models} & \multicolumn{2}{c}{Domain 1} & \multicolumn{2}{c}{Domain 2} \\ 
\cline{2-5}
          & DSC$\uparrow$   & ASD$\downarrow$   & DSC$\uparrow$   & ASD$\downarrow$  \\ \hline
Intra-domain & (80.06, 95.82) & (20.13, 7.53) & (73.13, 87.79) & (24.91, 18.75)  \\
\hline
DeepAll & (79.04\newcontent{$\pm$2.10}, 95.82\newcontent{$\pm$0.93}) & (20.32\newcontent{$\pm$2.14}, 7.63\newcontent{$\pm$1.55}) & (73.02\newcontent{$\pm$2.51}, 87.34\newcontent{$\pm$3.15}) & (24.99\newcontent{$\pm$3.67}, 18.70\newcontent{$\pm$1.44})  \\
BigAug~\cite{zhang_generalizing_2020} & (80.37\newcontent{$\pm$1.99}, 95.59\newcontent{$\pm$1.84}) & (19.50\newcontent{$\pm$1.59}, 7.75\newcontent{$\pm$1.43}) & (74.73\newcontent{$\pm$2.58}, 87.40\newcontent{$\pm$2.99}) & (22.64\newcontent{$\pm$3.28}, 18.89\newcontent{$\pm$1.39}) \\
SAML~\cite{liu_shape-aware_2020} & (81.03\newcontent{$\pm$1.91}, 95.74\newcontent{$\pm$1.62}) & (19.31\newcontent{$\pm$1.88}, 7.66\newcontent{$\pm$1.18}) & (76.61\newcontent{$\pm$1.99}, 87.29\newcontent{$\pm$2.79}) & (19.31\newcontent{$\pm$2.33}, 19.20\newcontent{$\pm$1.64})  \\
FedDG~\cite{liu_feddg_2021} & (81.66\newcontent{$\pm$1.37}, 95.47\newcontent{$\pm$1.55}) & (18.79\newcontent{$\pm$1.46}, 7.81\newcontent{$\pm$1.12}) & (76.31\newcontent{$\pm$2.09}, 86.34\newcontent{$\pm$3.26}) & (19.98\newcontent{$\pm$2.69}, 19.57\newcontent{$\pm$1.52})  \\
DoFE~\cite{wang_dofe_2020} & (\colorbox{cblue}{\textbf{81.95\newcontent{$\pm$1.02}}}, \colorbox{cgreen}{96.04\newcontent{$\pm$1.19}}) & (\colorbox{cblue}{\textbf{18.59\newcontent{$\pm$1.30}}}, \colorbox{cgreen}{7.05\newcontent{$\pm$0.90}}) & (\colorbox{cblue}{\textbf{78.31\newcontent{$\pm$1.75}}}, \colorbox{cgreen}{\textbf{89.20\newcontent{$\pm$1.72}}}) & (\colorbox{cblue}{\textbf{16.40\newcontent{$\pm$1.86}}}, \colorbox{cgreen}{\textbf{15.75\newcontent{$\pm$1.14}}})  \\ \hline
Ours (DCAC) & (\colorbox{cblue}{81.43\newcontent{$\pm$1.58}}, \colorbox{cgreen}{\textbf{96.54\newcontent{$\pm$0.84}}}) & (\colorbox{cblue}{19.20\newcontent{$\pm$1.79}}, \colorbox{cgreen}{\textbf{6.35\newcontent{$\pm$0.81}}}) & (\colorbox{cblue}{77.72\newcontent{$\pm$1.97}}, \colorbox{cgreen}{87.85\newcontent{$\pm$2.63}}) & (\colorbox{cblue}{17.15\newcontent{$\pm$2.30}}, \colorbox{cgreen}{18.28\newcontent{$\pm$1.33}})  \\ \hline \hline
\end{tabular}

\begin{tabular}{l|cccc|cc}
\hline \hline
\multirow{2}{*}{Models} & \multicolumn{2}{c}{Domain 3} & \multicolumn{2}{c}{Domain 4} & \multicolumn{2}{|c}{Average} \\ 
\cline{2-7}
          & DSC$\uparrow$   & ASD$\downarrow$   & DSC$\uparrow$   & ASD$\downarrow$   & DSC$\uparrow$   & ASD$\downarrow$  \\ \hline
Intra-domain & (83.80, 93.20) & (11.20, 9.64) & (84.46, 93.41) & (8.99, 7.51) & 86.46 & 13.58  \\
\hline
DeepAll & (82.26\newcontent{$\pm$1.58}, 91.37\newcontent{$\pm$1.80}) & (12.01\newcontent{$\pm$1.61}, 11.40\newcontent{$\pm$1.77}) & (84.85\newcontent{$\pm$2.96}, 92.27\newcontent{$\pm$2.82}) & (8.39\newcontent{$\pm$1.37}, 7.83\newcontent{$\pm$2.40}) & 85.75 & 13.91  \\
BigAug~\cite{zhang_generalizing_2020} & (85.39\newcontent{$\pm$1.42}, 92.04\newcontent{$\pm$1.51}) & (10.07\newcontent{$\pm$1.03}, 11.09\newcontent{$\pm$1.50}) & (86.47\newcontent{$\pm$1.47}, 93.05\newcontent{$\pm$1.57}) & (8.32\newcontent{$\pm$0.88}, 7.75\newcontent{$\pm$1.98}) & 86.88 & 13.25  \\
SAML~\cite{liu_shape-aware_2020} & (85.40\newcontent{$\pm$1.49}, \colorbox{cgreen}{93.92\newcontent{$\pm$0.93}}) & (\colorbox{cgreen}{9.99\newcontent{$\pm$0.68}}, \colorbox{cgreen}{8.62\newcontent{$\pm$0.66}}) & (86.06\newcontent{$\pm$2.51}, \colorbox{cgreen}{94.76\newcontent{$\pm$0.89}}) & (8.86\newcontent{$\pm$0.92}, \colorbox{cgreen}{5.90\newcontent{$\pm$0.87}}) & 87.60 & 12.36  \\
FedDG~\cite{liu_feddg_2021} & (85.23\newcontent{$\pm$1.10}, 93.36\newcontent{$\pm$0.99}) & (10.86\newcontent{$\pm$0.87}, 9.12\newcontent{$\pm$0.70}) & (85.27\newcontent{$\pm$2.90}, 94.68\newcontent{$\pm$1.12}) & (8.94\newcontent{$\pm$1.24}, 6.02\newcontent{$\pm$0.99}) & 87.29 & 12.64  \\
DoFE~\cite{wang_dofe_2020} & (\colorbox{cgreen}{85.51\newcontent{$\pm$0.94}}, 93.23\newcontent{$\pm$1.04}) & (10.06\newcontent{$\pm$1.26}, 9.76\newcontent{$\pm$1.22}) & (\colorbox{cgreen}{86.61\newcontent{$\pm$1.08}}, 94.28\newcontent{$\pm$1.07}) & (\colorbox{cgreen}{8.28\newcontent{$\pm$1.01}}, 6.99\newcontent{$\pm$1.16}) & 88.14 & 11.61  \\ \hline
Ours (DCAC) & (\colorbox{cgreen}{\textbf{86.80\newcontent{$\pm$0.76}}}, \colorbox{cgreen}{\textbf{94.28\newcontent{$\pm$0.84}}}) & (\colorbox{cgreen}{\textbf{9.14\newcontent{$\pm$0.62}}}, \colorbox{cgreen}{\textbf{8.11\newcontent{$\pm$0.56}}}) & (\colorbox{cgreen}{\textbf{87.68\newcontent{$\pm$0.85}}}, \colorbox{cgreen}{\textbf{95.40\newcontent{$\pm$0.62}}}) & (\colorbox{cgreen}{\textbf{7.12\newcontent{$\pm$0.72}}}, \colorbox{cgreen}{\textbf{5.20\newcontent{$\pm$0.68}}}) & \textbf{88.47} & \textbf{11.32}  \\ \hline \hline
\end{tabular}
\end{table*}

\begin{figure*}[]
    \centering
    \includegraphics[scale=0.43]{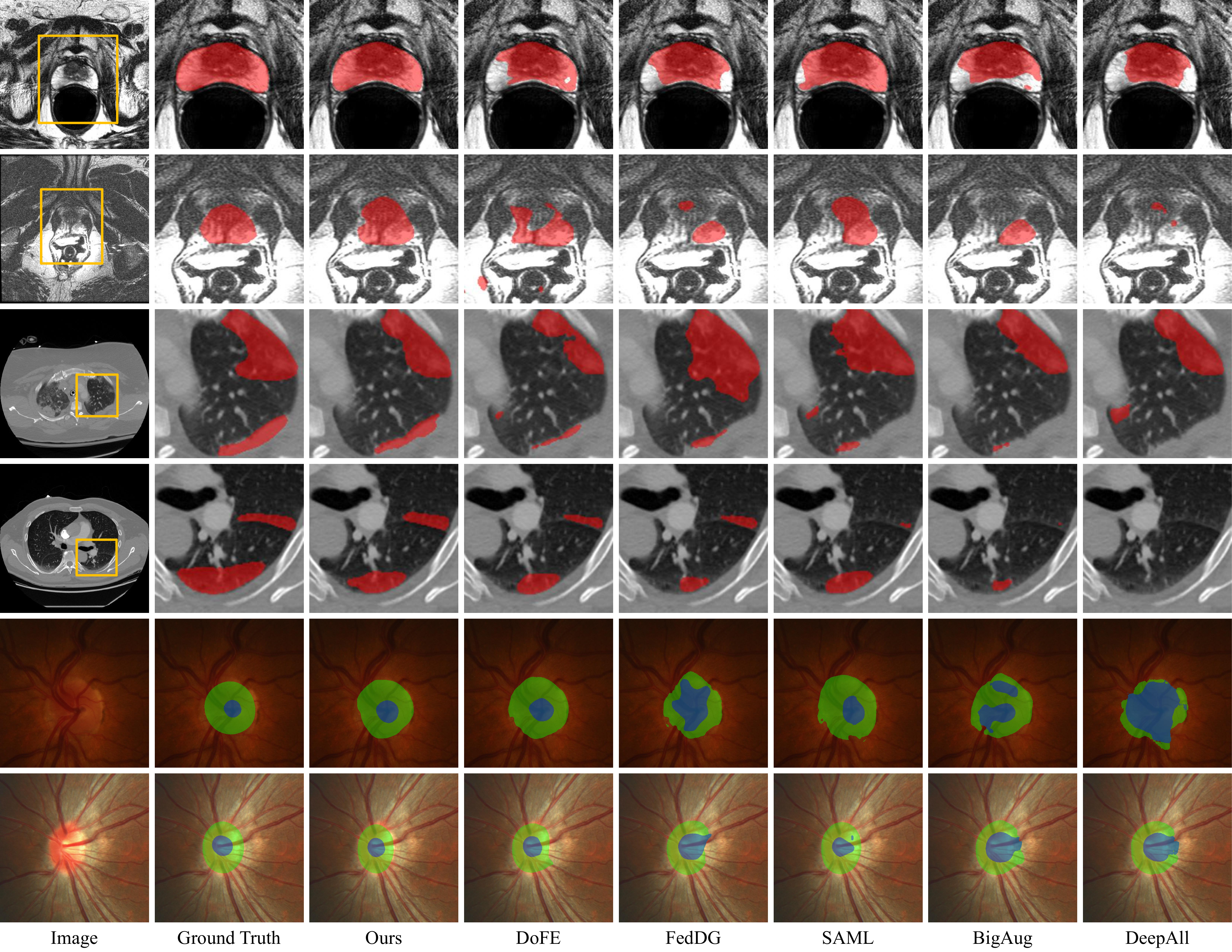}
    \caption{Visualization of the results predicted by ours (DCAC) and five competing methods on the three segmentation tasks, together with ground truth. Best viewed in color.}
    \label{fig:compare-visualization}
\end{figure*}

\section{Experiments}
We evaluated the proposed DCAC model against the baseline and state-of-the-art domain generalization models on three tasks, including prostate segmentation using MRI, COVID-19 lesion segmentation using CT, and OC/OD segmentation using color fundus image.
These tasks cover different image modalities and represent variable domain shifts in cross-domain medical image segmentation problems.

\subsection{Datasets}
Three datasets were used for this study.
For prostate segmentation, the dataset contains 116 T2-weighted MRI cases from six domains~\cite{liu_shape-aware_2020,bloch2015nci,lemaitre2015computer,litjens2014evaluation}. Following~\cite{liu_shape-aware_2020,liu_feddg_2021}, we preprocessed the MRI data and only preserved the slices with the prostate region for consistent and objective segmentation evaluation.
For COVID-19 lesion segmentation, the dataset consists of 120 RT-PCR positive CT scans with pixel-level lesion annotations, collected from the first multi-institutional, multi-national expert annotated COVID-19 image database~\cite{tsai_rsna_2021,clark2013cancer,ricord1a}.
For OC/OD segmentation, the dataset contains 789 cases for training and 281 cases for test, which are collected from four public fundus image datasets and have inconsistent statistical characteristics~\cite{wang_dofe_2020,sivaswamy2015comprehensive,fumero2011rim,orlando2020refuge}.
The statistics of three datasets were summarized in Table~\ref{tab:datasets}.

\subsection{Implementation Details}
The images in each segmentation task were normalized by subtracting the mean and dividing by the standard deviation.
To make a compromise between the network complexity and input image size, the mini-batch size was set to 32 for 2D prostate segmentation with a patch size of $256\times256$, set to 16 for 2D OC/OD segmentation with a patch size of $512\times 512$, and set to 2 for 3D COVID-19 lesion segmentation with a patch size of $128\times196\times196$.
To expand the training set, several data augmentation techniques were used, including random cropping, rotation, scaling, flipping, adding Gaussian noise, and elastic deformation.
The SGD algorithm with a momentum of 0.99 was adopted as the optimizer.
The initial learning rate $lr_0$ was set to 0.01 and decayed according to the polynomial rule $lr = lr_0 \times (1-t/T)^{0.9}$, where $t$ is the current epoch and $T$ is the maximum epoch.
The maximum epoch $T$ was set to 200 for 2D prostate segmentation, 500 for 2D OC/OD segmentation, and 1000 for 3D COVID-19 lesion segmentation.
Our DCAC was implemented using the PyTorch framework on a workstation with a NVIDIA 2080Ti GPU.

\begin{table*}[]
\centering
\caption{Performance of DeepAll, our DCAC model, and its six variants in prostate segmentation.}
\label{tab:dac/cac}
\renewcommand{\arraystretch}{1} 
\setlength\tabcolsep{3pt}
\begin{tabular}{l|cccccccccccc|cc}
\hline \hline
\multirow{2}{*}{Models} & \multicolumn{2}{c}{Domain 1} & \multicolumn{2}{c}{Domain 2} & \multicolumn{2}{c}{Domain 3} & \multicolumn{2}{c}{Domain 4} & \multicolumn{2}{c}{Domain 5} & \multicolumn{2}{c}{Domain 6} & \multicolumn{2}{|c}{Average} \\ 
\cline{2-15}
          & DSC$\uparrow$   & ASD$\downarrow$   & DSC$\uparrow$   & ASD$\downarrow$   & DSC$\uparrow$   & ASD$\downarrow$   & DSC$\uparrow$   & ASD$\downarrow$   & DSC$\uparrow$   & ASD$\downarrow$   & DSC$\uparrow$   & ASD$\downarrow$   & DSC$\uparrow$   & ASD$\downarrow$   \\ \hline
DeepAll & 89.16 & 2.09 & 87.31 & 1.27 & 74.12 & 3.02 & 88.85 & 2.36 & 83.22 & 3.51 & 88.39 & 1.67 & 85.18 & 2.32 \\
D-CAC & 91.24 & 1.37 & 89.94 & 0.92 & 86.72 & 1.67 & 89.23 & 1.34 & 79.51 & 3.54 & 89.90 & 0.96 & 87.74 & 1.70 \\
Ours w/o DAC & 91.13 & 1.12 & 89.62 & 1.01 & 84.75 & 2.17 & 89.31 & 1.48 & 80.79 & 2.11 & 89.93 & 0.93 & 87.59 & 1.47 \\
Ours w/o CAC & 91.69 & 1.01 & 89.96 & 0.97 & 85.27 & 1.89 & 89.19 & 1.33 & 78.44 & 2.35 & 90.65 & 0.90 & 87.53 & 1.41 \\
\hline \hline
CDAC & 91.74 & 1.11 & 90.72 & 0.90 & 86.05 & 2.15 & 89.18 & 1.52 & 83.27 & 2.02 & 90.21 & 0.94 & 88.53 & 1.44  \\
DC$^{(p)}$AC & 90.02 & 1.61 & 89.78 & 0.96 & 85.44 & 1.94 & 88.28 & 2.04 & 81.37 & 2.34 & 90.41 & 0.88 & 87.55 & 1.63  \\
DCAC-NG & 90.67 & 1.29 & 88.12 & 1.04 & 78.58 & 2.87 & 88.31 & 1.41 & 81.18 & 1.97 & 89.69 & 1.01 & 86.09 & 1.60  \\
\hline \hline
Ours (DCAC) & 91.76 & 0.98 & 90.51 & 0.89 & 86.30 & 1.77 & 89.13 & 1.53 & 83.39 & 2.46 & 90.56 & 0.85 & 88.61 & 1.41 \\
\hline \hline
\end{tabular}
\end{table*}

\begin{figure*}[]
    \centering
    \includegraphics[scale=0.5]{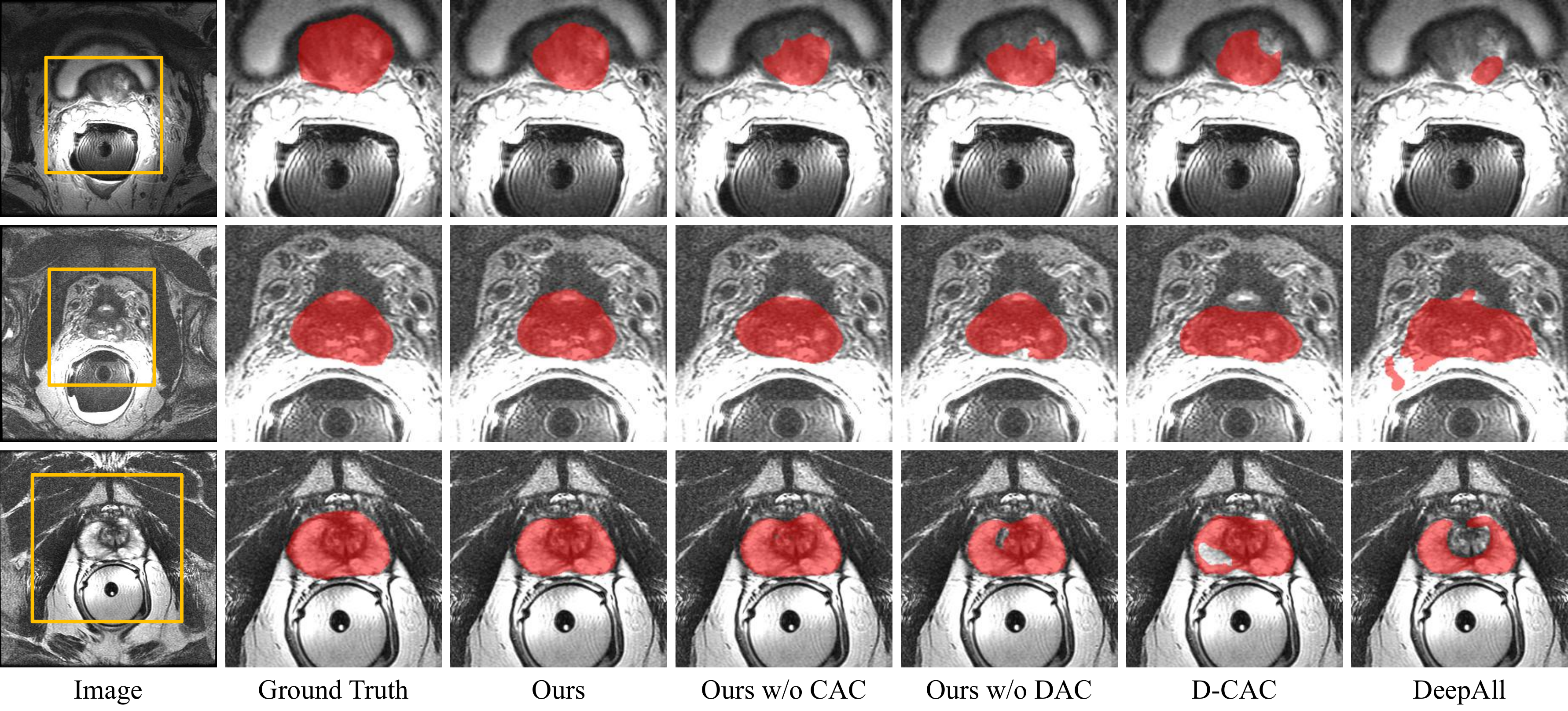}
    \caption{Visualization of one slice from each of three prostate MRI scans, corresponding segmentation ground truth, and the results obtained by applying our DCAC, three variants of DCAC, and DeepAll.}
    \label{fig:ablation-visualization}
\end{figure*}

\begin{figure}[]
  \centering
  \includegraphics[scale=0.38]{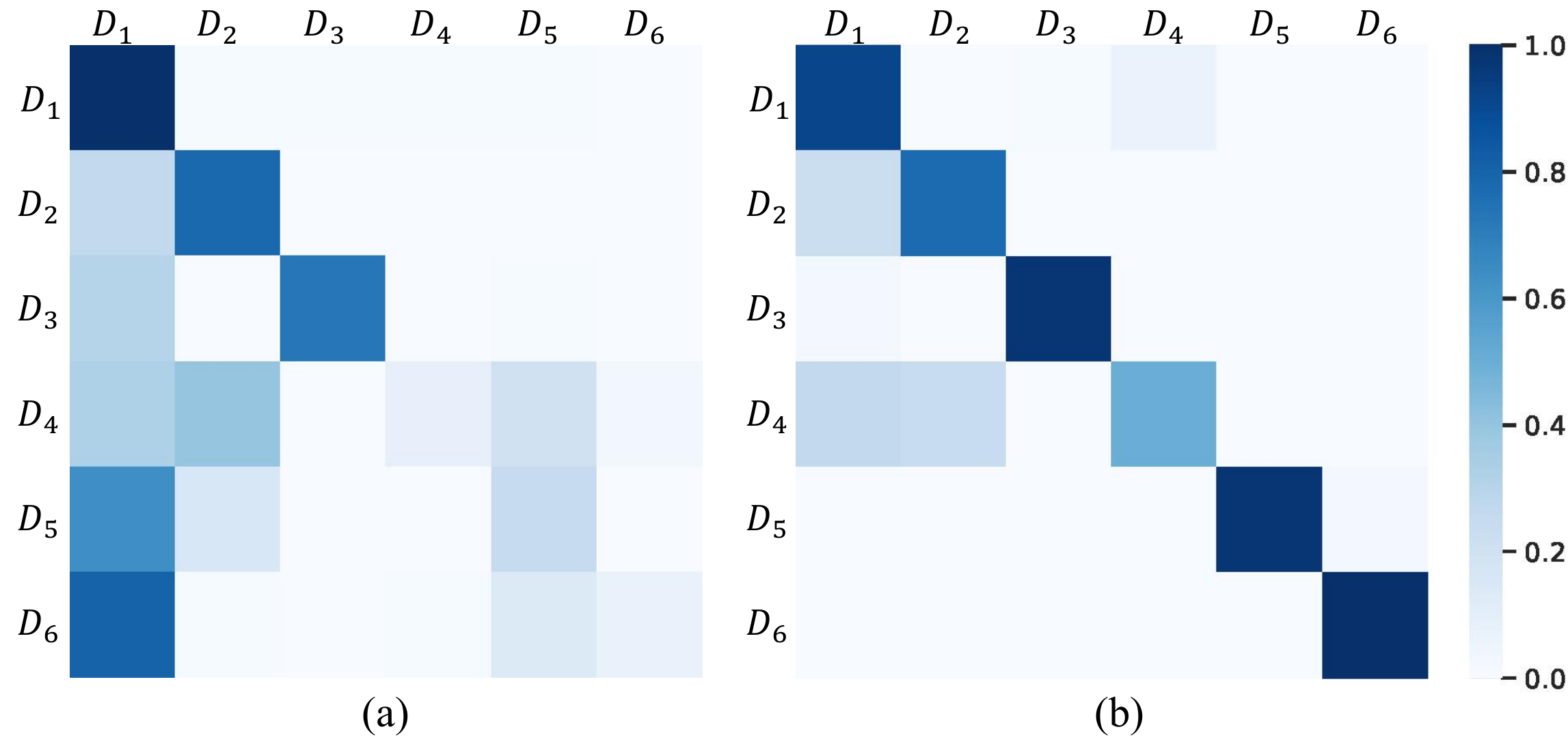}
  \caption{Confusion matrices of domain classification achieved by using (a) single-scale global features or (b) multi-scale global features as input of domain predictor. In each confusion matrix, the rows and columns represent true and predicted domains, respectively, and each element $(i, j)$ represents the probability of predicting domain $D_i$ as $D_j$. Particularly, a diagonal element represents the true positive rate of predicting the corresponding domain.
}
  \label{fig:confusion-matrix}
\end{figure}

\subsection{Comparative Experiments and Analysis}
We compared the proposed DCAC model with the `Intra-domain' setting, `DeepAll' baseline, and four domain generalization methods including 
(1) a data-augmentation based method called BigAug~\cite{zhang_generalizing_2020},
(2) two meta-learning methods called SAML~\cite{liu_shape-aware_2020} and FedDG~\cite{liu_feddg_2021}, and
(3) a domain-invariant feature learning approach called DoFE~\cite{wang_dofe_2020}.
Under the `Intra-domain' setting, training and test data are from the same domain and the three-fold cross-validation is used. Whereas under the `DeepAll' setting, the model is trained on the data aggregated from all source domains and tested directly on the unseen target domain.
\newcontent{
Note that the backbone, \textit{i.e.}, nnUNet, was kept as the same for all these methods in all experiments unless otherwise indicated.}
For each segmentation task, the leave-one-domain-out strategy was used to evaluate the performance of each domain generalization method, \textit{i.e.}, training on $K-1$ source domains and evaluating on the left unseen target domain. Each domain is chosen as the target domain in turn.
The segmentation performance was measured by the Dice Similarity Coefficient (DSC) and Average Surface Distance (ASD).
The DSC (\%) and ASD (pixel) characterize the accuracy of predicted masks and boundaries, respectively.

\subsubsection{Comparative results in prostate segmentation}
Table~\ref{tab:prostate} gives the DSC and ASD values obtained by our DCAC model and six segmentation models in each target domain and the average performance over six domains.
As expected, the performance of DeepAll seems to be worse on average than that of Intra-domain, due to the distribution discrepancy between the source (training) data and target (test) data.
Meanwhile, it shows that the augmentation-based method BigAug performs worse than meta-learning-based methods (\textit{i.e.}, SAML and FedDG), indicating that simply augmenting training data is insufficient to simulate the data distribution of the target domain.
It also shows that DoFE is superior to FedDG but slightly inferior to SAML, suggesting that the domain-invariant feature learning approach (\textit{i.e.}, DoFE) can disentangle domain-sensitive features, but it can hardly adapt to different domain discrepancies automatically. 
More importantly, it reveals that the proposed DCAC mode not only beats Intra-domain and DeepAll but also outperforms four state-of-the-art domain generalization methods. We believe the superior performance can be attributed to the fact that, with dynamic convolution, our model is capable of adapting to both the predicted domain code and extracted global features of the input image. 

\subsubsection{Comparative results in COVID-19 lesion segmentation and OC/OD segmentation}
The segmentation performance of our DCAC model and six segmentation models on the COVID-19 lesion segmentation task and OC/OD segmentation task was given in Table~\ref{tab:covid19} and Table~\ref{tab:oc/od}, respectively.
In COVID-19 lesion segmentation, the average DSC of Intra-domain is surprisingly worse than that of DeepAll. A possible reason is that the amount of training data in a single domain (see Table~\ref{tab:datasets}) is far from sufficient for training a DCNN model, leading to serious over-fitting of the small training dataset.
By contrast, aggregating the data in multiple domains can benefit model training and thus results in improved performance.
Meanwhile, it seems that the domain-invariant feature learning method is relatively better than meta-learning methods in both experiments, indicating the sensitivity of meta-learning-based methods to the number of source domains. When there are less source domains, the diversity of the generalization gap simulated by meta-learning is highly restricted.
Comparing to these methods, our model is less susceptible to the number of source domains and achieves stable performance gain on both segmentation tasks. This observation is consistent with what we observed in Table~\ref{tab:prostate}.

\subsubsection{Visualization results of DCAC and other competing methods}
We visualized the segmentation results of our DCAC and four competing domain generalization methods in Fig.~\ref{fig:compare-visualization}. We also displayed the results of DeepAll and ground truth for reference. It shows that the segmentation results produced by our DCAC model are the most similar to the ground truth over all three segmentation tasks, which confirms the effectiveness of our DCAC model against the state-of-the-art in the three generalizable medical image segmentation benchmarks with different imaging modalities.

\newcontent{
We calculated the standard deviation for every performance} \newcontent{metric and performed the independent two sample t-test between our result and the result of best competing model in each domain, as shown in Table~\ref{tab:prostate}, Table~\ref{tab:covid19}, and Table~\ref{tab:oc/od}. If our result is statistically different from the competing one at the $p = 0.05$ level, the corresponding cell in the table is highlighted in green; otherwise, the cell is highlighted in blue. It shows that our DCAC model outperforms mostly the best competing model and the performance gain is statistically significant in most cases. Nevertheless, when DCAC underperforms the best competing model, the results achieved in most of such cases are not statistically different.
}

\section{Discussion}

The prostate segmentation task was chosen as a case study, and ablation studies were conducted on this task to investigate the effectiveness of newly designed DAC and CAC modules and the domain-discriminatory ability of extracted features. 

\subsection{Ablation Analysis}

In this work, we designed the DAC module and CAC module to make our model capable of adapting to the unseen test domain and test image, respectively.
To evaluate the contributions of these two modules, we compared our model with its variant that uses only one module. 
Meanwhile, we changed the order between DAC and CAC for comparison, denoted as CDAC. We also compared a variant, denoted by D-CAC, that uses the concatenation of domain code and global image features to generate one and only one unified dynamic head.
The performance of DeepAll (Baseline), our DCAC model, and its variants was given in Table~\ref{tab:dac/cac}.
It shows that CDAC achieves similar performance to our DCAC, and both of them outperform not only D-CAC but also the variant without either DAC or CAC. The results confirm that either DAC or CAC contributes to the final results and the two-dynamic-head strategy is superior to the unified dynamic head.

We visualized the segmentation results of DCAC and three variants in Fig.~\ref{fig:ablation-visualization}. We also displayed the results of DeepAll and ground truth for reference. It shows that our DCAC model can produce more accurate segmentation results of unseen test images, particularly in the boundary region.

\subsection{Domain-discriminatory Power of Extracted Features}
In our DCAC model, the DAC module relies heavily on the domain code $\mathscr{D}^p$ predicted based on image features. 
To predict $\mathscr{D}^p$ accurately, the features should have sufficient domain-discriminatory power. Instead of using the single-scale global feature produced by the last encoder block, we chose the multi-scale global features extracted by the encoder at all scales for our study (see Fig.~\ref{fig:overview}).
To verify the superiority of our multi-scale features, we compared the domain classification accuracy achieved by using each of these two types of features. 
The obtained confusion matrices were visualized in Fig.~\ref{fig:confusion-matrix}.
It shows that using multi-scale features can produce more accurate domain classification than using single-scale features, suggesting that the multi-scale feature maps produced by encoder blocks contain domain-specific information and can be used to generate the domain code.

\subsection{Analysis of Complexity}

\begin{table}[]
\centering
\caption{Number of parameters, GFLOPs, model size, and training duration time cost of different models in prostate segmentation.}
\label{tab:complexity}
\setlength\tabcolsep{4pt}
\renewcommand{\arraystretch}{1.2} 
\begin{tabular}{l|cccc}
\hline
\hline
Models & \makecell[c]{\#Parameters\\\newcontent{($\times10^6$)}} & GFLOPs & \makecell[c]{Model\\Size (MB)} & \makecell[c]{Training\\Time Cost (Hours)} \\ \hline
DoFE~\cite{wang_dofe_2020} & 30.1 & 32.7 & 145.4 & 7.1 \\ \hline
SAML~\cite{liu_shape-aware_2020} & 30.0 & 32.4 & 119.9 & 11.8 \\ \hline
Ours (DCAC) & 30.1 & 32.5 & 120.5 & 6.2 \\ \hline \hline
\end{tabular}
\end{table}

Besides the segmentation backbone, the proposed DCAC model also contains a domain predictor, two controllers, and two dynamic convolutional heads, which, fortunately, all have lightweight structures. Therefore, apart from the parameters in the backbone, DCAC just has a few extra parameters and consumes a little extra training time.
We chose the prostate segmentation task as a case study and listed the number of parameters, GFLOPs, model size, and training time cost of our DCAC model and two state-of-the-art domain generalization methods (\textit{i.e.}, DoFE and SAML) in Table~\ref{tab:complexity}.
\newcontent{Note that the parameters of the entire model (including the backbone) were counted for all methods, and the backbone was also taken into} \newcontent{consideration when calculating GFLOPs and the model size.}
It shows that, although three models have a similar number of parameters and GFLOPs, the size of DCAC model is significantly smaller than DoFE, since DoFE uses an additional domain knowledge pool to store domain prior knowledge for domain-sensitive feature matching during inference, but our DCAC achieves this using the light-weighted dynamic convolutions.
\newcontent{Moreover, our DCAC has much less training time cost than DoFE and SAML. The extremely high time cost of SAML can be attributed to the fact that SAML relies heavily on meta-train and meta-test to update model parameters in each iteration, which is time-consuming. As for DoFE, it requires to update the domain knowledge pool and perform feature embedding during each training step.} 
In summary, our results indicate that, comparing to DoFE and SAML, the proposed DCAC model is able to produce more accurate segmentation results with less spatial and computational complexity.

\subsection{Applying to Other Backbone}

\begin{figure}[]
    \centering
    \includegraphics[scale=0.6]{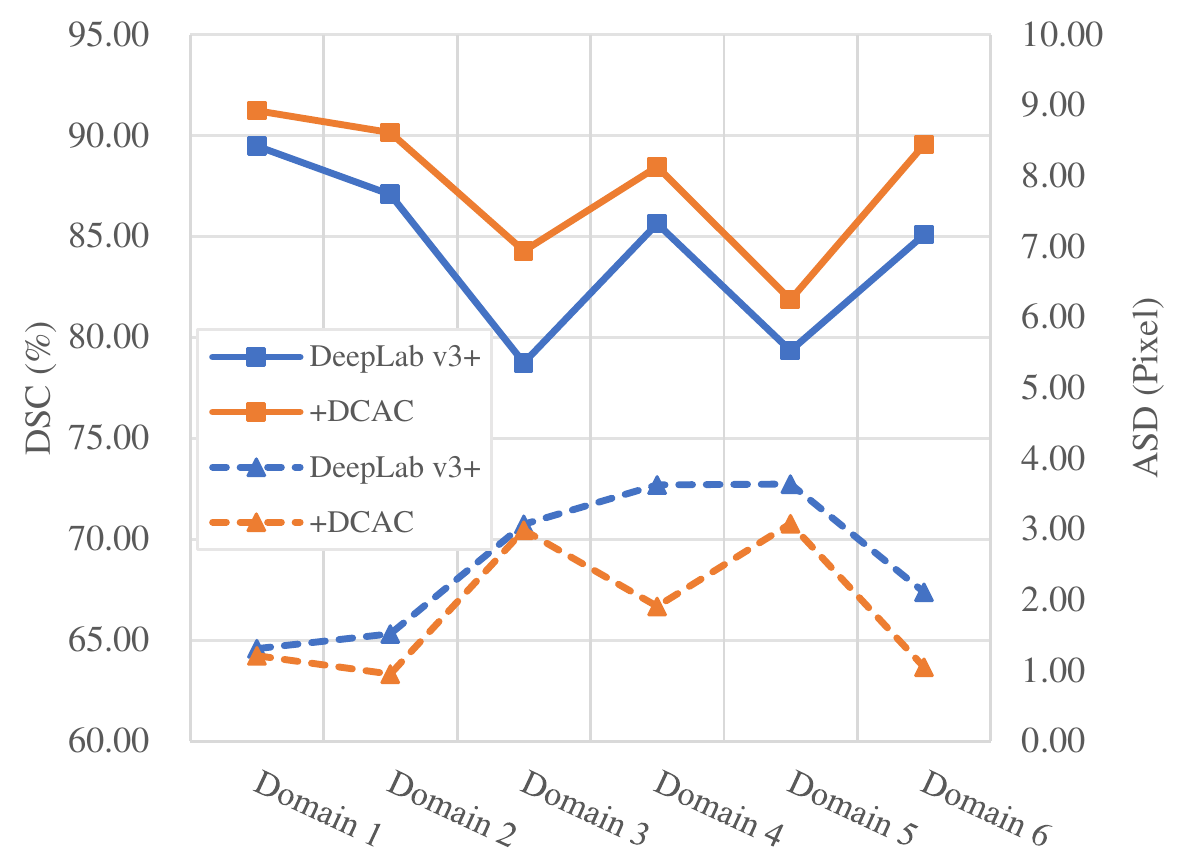}
    \caption{Performance of DeepLab V3+ with and without DCAC on prostate segmentation task. The values in square represent DSC, and the values in triangle denote ASD.}
    \label{fig:modular}
\end{figure}

The DCAC is proposed following a modular design and can be incorporated into other encoder-decoder backbones to improve their performance. To validate this, we adopted DeepLab V3+~\cite{chen2018encoder} as the segmentation backbone and incorporated DCAC into it by adding a domain predictor and two dynamic heads.
We carried out the prostate segmentation task again and compared the performance of DeepLab V3+ with and without DCAC on six domains in Fig.~\ref{fig:modular}. It reveals that plugging DCAC into DeepLab V3+ results in performance gains on all domains, evidenced by the consistent increase of DSC and decrease of ASD. 
On average, the mean DSC improves from 84.23\% to 87.59\%, and the mean ASD drops from 2.55 to 1.87. The results are consistent with those reported in Table~\ref{tab:prostate}, confirming the usability and effectiveness of our DCAC again.

\subsection{Generalization Analysis of DAC and CAC}

\begin{figure}[]
    \centering
    \includegraphics[scale=0.21]{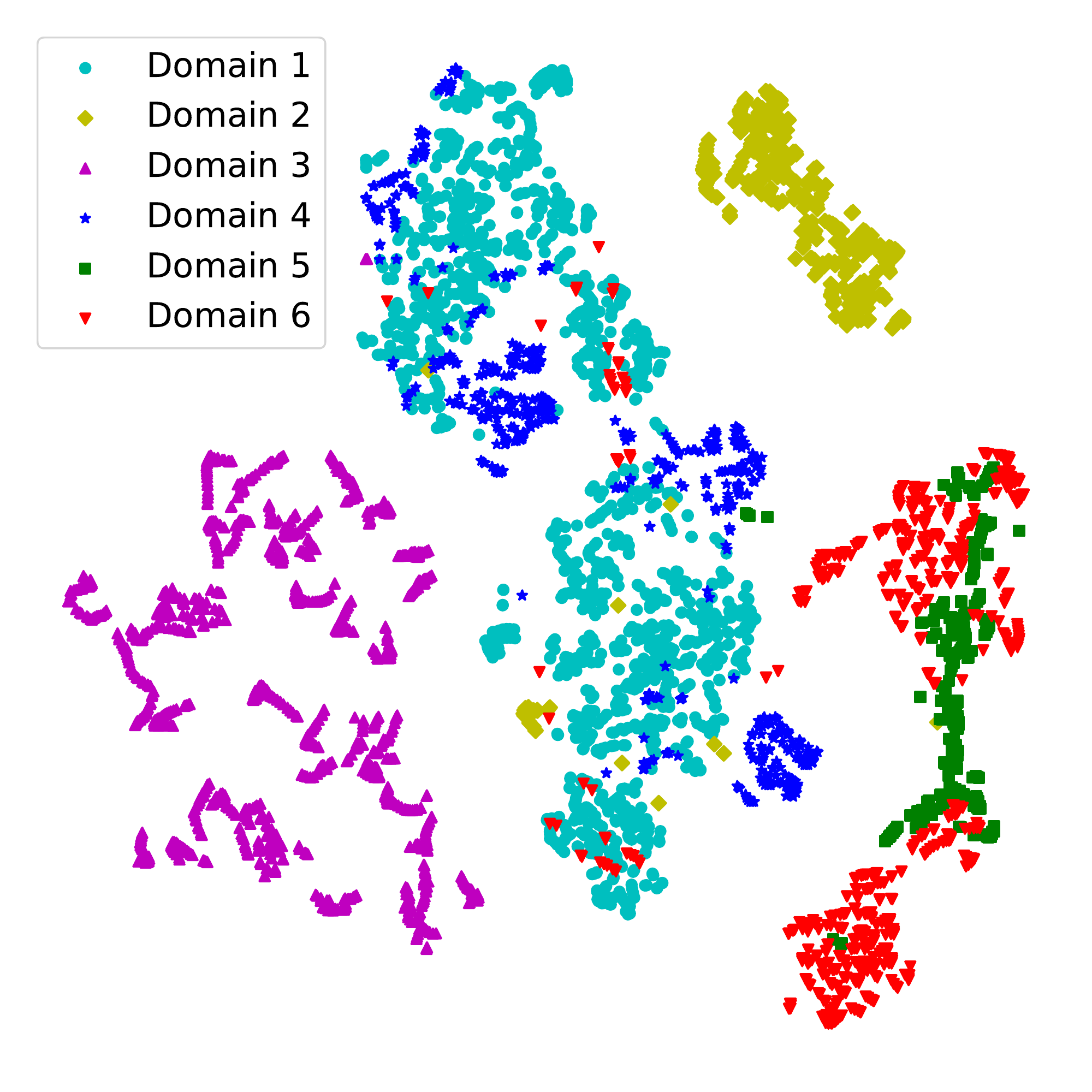}
    \hspace{1pt}
    \includegraphics[scale=0.21]{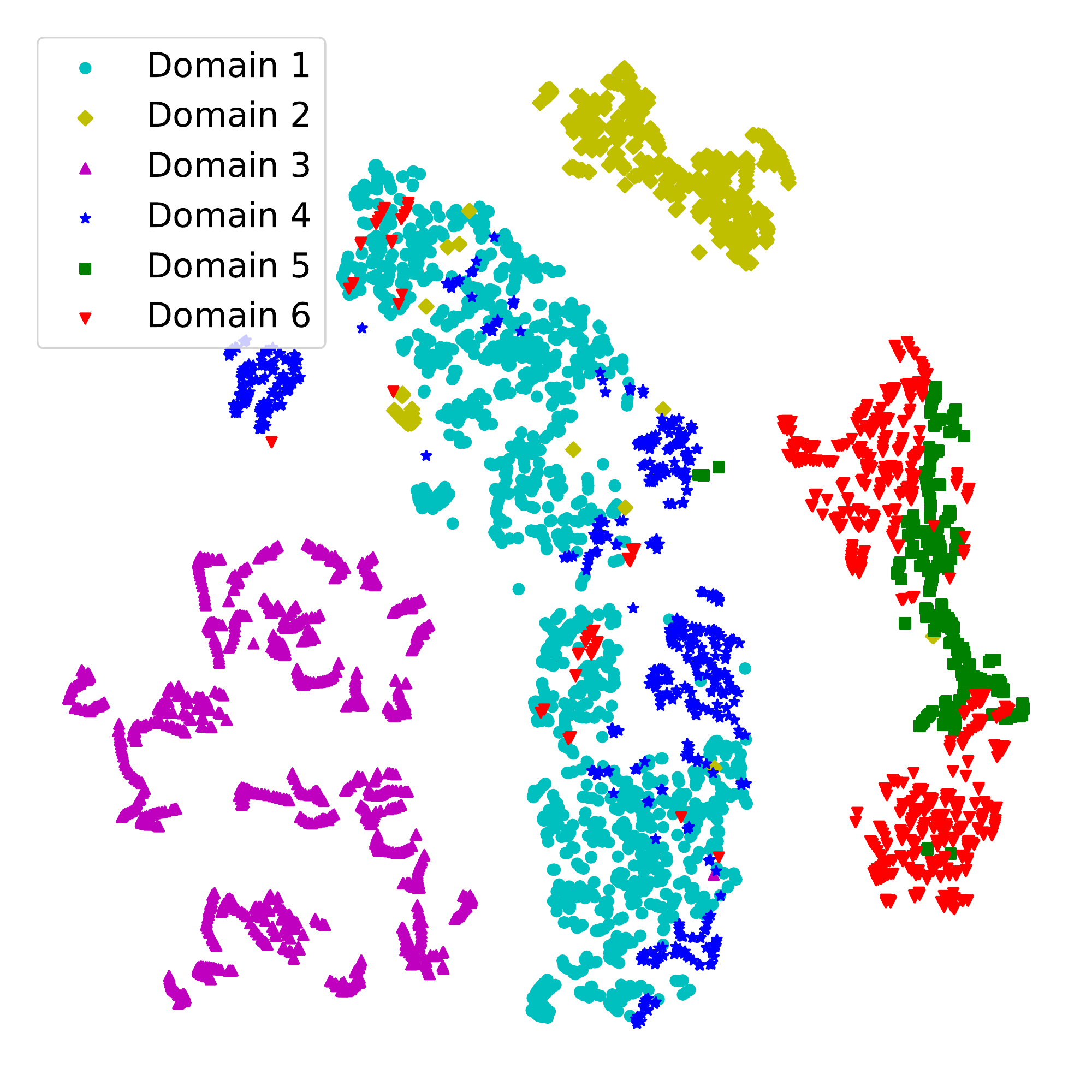}
    \caption{Visualization of feature maps of input image (right) with and (left) without perturbations in 2D using t-SNE. The points in different colors represent the images from different domains.}
    \label{fig:generalization_dac}
\end{figure}

To make the proposed DCAC model generalizable to unseen target domains, the domain adaptive head itself should be generalizable. To validate this, we chose the model trained using Domain 1 $\sim$ Domain 5 on the prostate segmentation task as a case study. The feature map of each input image (including both training and test cases) produced by the DAC module, denoted by ${Conv}_O^1(f_D^1)*\omega_d$, is visualized in 2D using t-SNE (see the left part of Fig.~\ref{fig:generalization_dac}). In the meantime, we added the white Gaussian noises with a standard deviation of 0.2 as perturbations to each input image and visualized the feature map produced by DAC in the right part of Fig.~\ref{fig:generalization_dac}. 
It shows that adding perturbations to input images leads to little impact on the topology of the DAC features. Therefore, the feature extraction performed by the DAC module is robust to input perturbations, suggesting that the DAC module does not over-fit the data from source domains.

\begin{figure}[]
    \centering
    \includegraphics[scale=0.21]{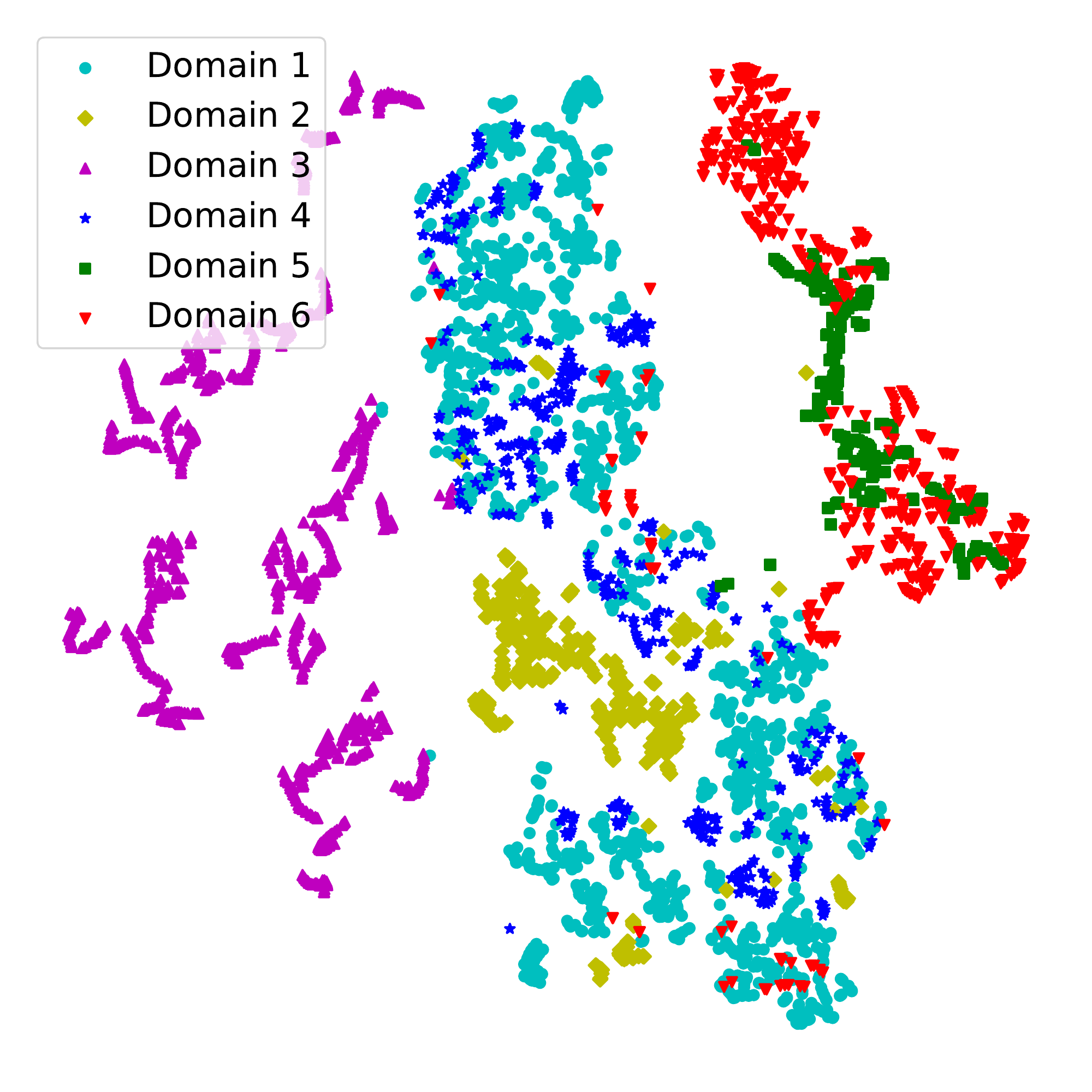}
    \hspace{1pt}
    \includegraphics[scale=0.21]{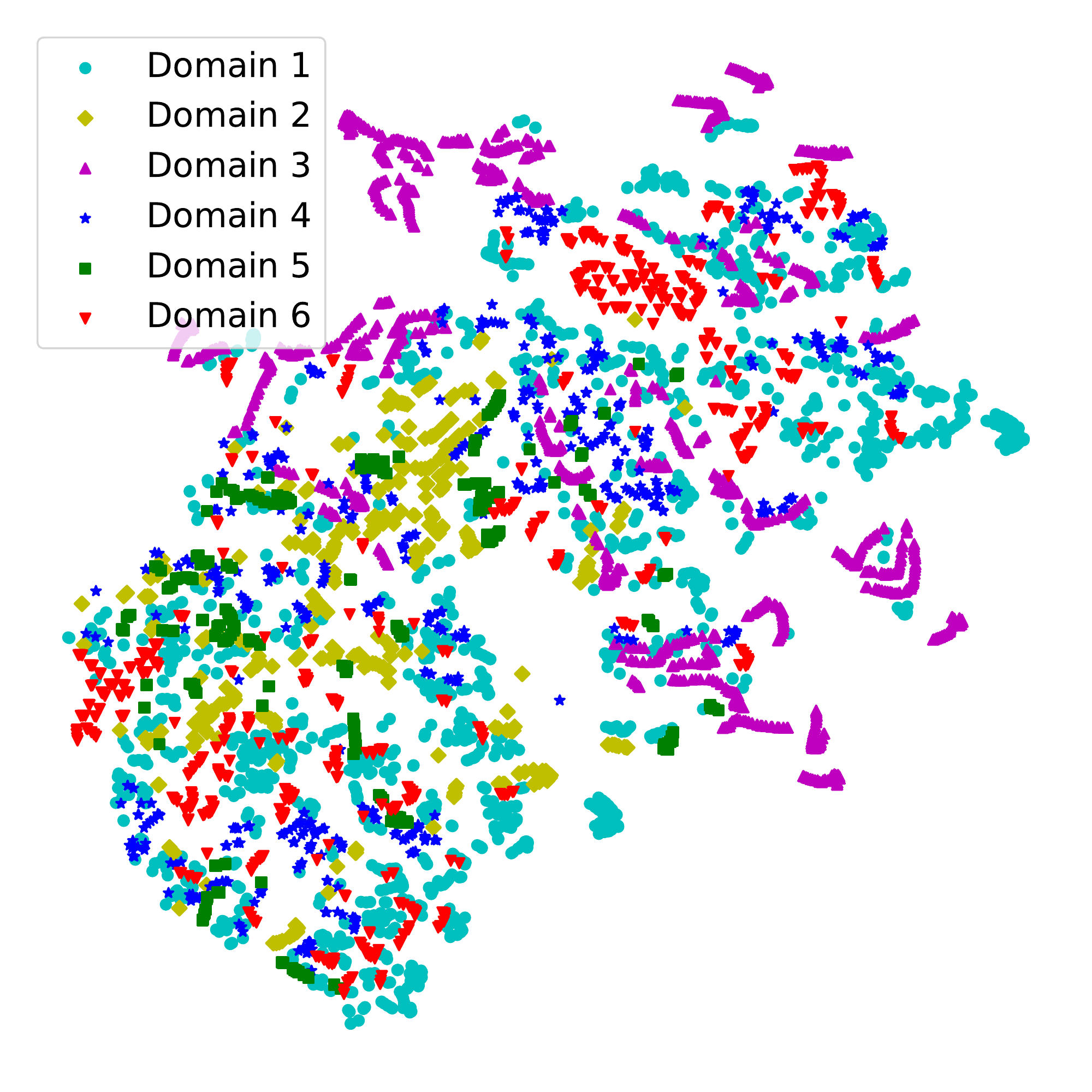}
    \caption{t-SNE visualization of features before (left) and after (right) being processed by the DAC head. The points in different colors represent images from different domains.}
    \label{fig:effect_dac}
\end{figure}

To demonstrate that the DAC convolutions $\omega_d$ can filter domain-specific features, we also visualized the image features before and after being filtered by $\omega_d$ in 2D using t-SNE (see Fig.~\ref{fig:effect_dac}).
It shows that, after being filtered by $\omega_d$, the image features from different domains, which previously can be largely separated from each other, become indistinguishable. It indicates that the DAC module can effectively filter out domain-specific features.

The dynamic convolutions in the CAC module, whose parameters $\omega_c$ are dynamically produced based on the global features of each input image, aim to adapt our DCAC model to the input. 
To validate the effectiveness of these convolutions, we shuffled the $\omega_c$ generated from different input images and evaluated the impact of such perturbations on the segmentation performance.
The segmentation performance of our DCAC model with (denoted by DC$^{(p)}$AC) or without $\omega_c$ perturbations on the prostate dataset was given in Table~\ref{tab:dac/cac}.
It shows that mutating $\omega_c$ deteriorates the performance of our DCAC model on each unseen target domain, decreasing the average DSC from 88.61\% to 87.55\%.
It confirms that the $\omega_c$ estimated by the content-aware controller is suitable for each input image.

\subsection{Gradient Truncation in Domain Predictor}
Besides image segmentation, the proposed DCAC model also performs domain classification using the domain predictor. These two tasks share the features extracted from the same encoder. For the segmentation task, the extracted features should be domain-insensitive so that the segmentation performance would be less affected by the domain discrepancy. Whereas the classification task requires the extracted features to have domain-discriminatory power.
To make a compromise between the domain classification accuracy and the segmentation performance, we adopt the gradients truncation strategy in the domain predictor. Thus, only the parameters in the fully-connect layer in the domain predictor can be optimized for domain classification.
It can be observed from Fig.~\ref{fig:confusion-matrix} that the domain attributions can still be largely discriminated when adopting the gradient truncation strategy. We also analyzed the segmentation performance when using (DCAC) and not using (DCAC-NG) gradients truncation.
The quantitative results were shown in Table~\ref{tab:dac/cac}. 
It reveals that the overall segmentation performance on the unseen target domains can be dramatically decreased when the shared encoder is optimized for both domain classification and segmentation. It confirms the benefit brought by the gradient truncation strategy.

\section{Conclusion}
This paper proposes a multi-source domain generalization model called DCAC, which uses two dynamic convolutional heads.
One dynamic head is conditioned on the predicted domain code of the input to make the DCAC model adapt to the target domain, while the other dynamic head is conditioned on global image features to make the model adapt to the input image.
Our results on the prostate segmentation, COVID-19 lesion segmentation, and OC/OD segmentation tasks suggest that, after training on the data from multiple source domains, the proposed DCAC model can generalize well on an unseen target domain, achieving improved average performance over the baseline and four state-of-the-art domain generalization methods.

However, the proposed DCAC model still has two limitations. First, it is designed for multi-source domain generalization, and therefore cannot be directly applied to single-source domain generalization~\cite{chen_cooperative_2021,ouyang_causality-inspired_2021}. In other words, it requires training data from multiple source domains. 
\newcontent{Nevertheless, we believe multi-source domain generalization is a promising research direction orthogonal to existing single-source domain generalization methods.}
Second, the performance gain achieved by our DCAC on CT images is less than that on MR and fundus images (see Table~\ref{tab:prostate}, Table~\ref{tab:covid19}, and Table~\ref{tab:oc/od}). 
It can be attributed to the fact that CT images for the same phase generally have more consistent image quality, whereas MR and fundus images hold more vendor-specific variations~\cite{zhang_generalizing_2020}.
In our future work, we will extend the proposed DCAC model to multi-source, multi-modality, and multi-task scenarios, aiming to provide a large-scale pre-trained segmentation model for various downstream medical image segmentation tasks.

\section{Acknowledgments}
We acknowledge the RSNA and Society of Thoracic Radiology (STR), the European Society of Medical Imaging Informatics, the American College of Radiology, and the American Association of Physicists in Medicine, and their critical role in the creation of the free publicly available RICORD dataset used for this study.
We also appreciate the efforts devoted by the authors of ~\cite{liu_shape-aware_2020} and ~\cite{wang_dofe_2020} to collect and share the prostate MR and fundus imaging data for comparing generalizable medical image segmentation algorithms.

\bibliographystyle{IEEEtran}
\bibliography{reference}

\begin{thebibliography}{10}
\providecommand{\url}[1]{#1}
\csname url@samestyle\endcsname
\providecommand{\newblock}{\relax}
\providecommand{\bibinfo}[2]{#2}
\providecommand{\BIBentrySTDinterwordspacing}{\spaceskip=0pt\relax}
\providecommand{\BIBentryALTinterwordstretchfactor}{4}
\providecommand{\BIBentryALTinterwordspacing}{\spaceskip=\fontdimen2\font plus
\BIBentryALTinterwordstretchfactor\fontdimen3\font minus
  \fontdimen4\font\relax}
\providecommand{\BIBforeignlanguage}[2]{{%
\expandafter\ifx\csname l@#1\endcsname\relax
\typeout{** WARNING: IEEEtran.bst: No hyphenation pattern has been}%
\typeout{** loaded for the language `#1'. Using the pattern for}%
\typeout{** the default language instead.}%
\else
\language=\csname l@#1\endcsname
\fi
#2}}
\providecommand{\BIBdecl}{\relax}
\BIBdecl

\bibitem{LITJENS201760}
G.~Litjens, T.~Kooi, B.~E. Bejnordi, A.~A.~A. Setio, F.~Ciompi, M.~Ghafoorian,
  J.~A. {van der Laak}, B.~{van Ginneken}, and C.~I. Sánchez, ``A survey on
  deep learning in medical image analysis,'' \emph{Medical Image Analysis},
  vol.~42, pp. 60--88, 2017.

\bibitem{xie_survey_2021}
X.~Xie, J.~Niu, X.~Liu, Z.~Chen, S.~Tang, and S.~Yu,
  ``\BIBforeignlanguage{en}{A survey on incorporating domain knowledge into
  deep learning for medical image analysis},''
  \emph{\BIBforeignlanguage{en}{Medical Image Analysis}}, vol.~69, p. 101985,
  Apr. 2021.

\bibitem{falk_u-net_2019}
T.~Falk, D.~Mai, R.~Bensch, {\"O}.~{\c{C}}i{\c{c}}ek, A.~Abdulkadir,
  Y.~Marrakchi, A.~B{\"o}hm, J.~Deubner, Z.~J{\"a}ckel, K.~Seiwald
  \emph{et~al.}, ``\BIBforeignlanguage{en}{U-{Net}: Deep learning for cell
  counting, detection, and morphometry},'' \emph{\BIBforeignlanguage{en}{Nature
  Methods}}, vol.~16, no.~1, pp. 67--70, Jan. 2019.

\bibitem{zhou_unet_2020}
Z.~Zhou, M.~M.~R. Siddiquee, N.~Tajbakhsh, and J.~Liang, ``{UNet}++:
  {Redesigning} {Skip} {Connections} to {Exploit} {Multiscale} {Features} in
  {Image} {Segmentation},'' \emph{IEEE Transactions on Medical Imaging},
  vol.~39, no.~6, pp. 1856--1867, Jun. 2020.

\bibitem{isensee_nnu-net_2020}
F.~Isensee, P.~F. Jaeger, S.~A.~A. Kohl, J.~Petersen, and K.~H. Maier-Hein,
  ``\BIBforeignlanguage{en}{{nnU}-{Net}: A self-configuring method for deep
  learning-based biomedical image segmentation},''
  \emph{\BIBforeignlanguage{en}{Nature Methods}}, vol.~18, no.~2, pp. 203--211,
  Dec. 2020.

\bibitem{guo2021learning}
R.~Guo, M.~Pagnucco, and Y.~Song, ``Learning with noise: Mask-guided attention
  model for weakly supervised nuclei segmentation,'' in \emph{Medical Image
  Computing and Computer Assisted Intervention (MICCAI)}.\hskip 1em plus 0.5em
  minus 0.4em\relax Springer, 2021, pp. 461--470.

\bibitem{wang2021deep}
X.~Wang, H.~Chen, H.~Xiang, H.~Lin, X.~Lin, and P.-A. Heng, ``Deep virtual
  adversarial self-training with consistency regularization for semi-supervised
  medical image classification,'' \emph{Medical Image Analysis}, vol.~70, p.
  102010, 2021.

\bibitem{huang2020noise}
Y.~Huang, W.~Xia, Z.~Lu, Y.~Liu, H.~Chen, J.~Zhou, L.~Fang, and Y.~Zhang,
  ``Noise-powered disentangled representation for unsupervised speckle
  reduction of optical coherence tomography images,'' \emph{IEEE Transactions
  on Medical Imaging}, 2020.

\bibitem{yang2021towards}
Y.~Yang, J.~Chen, R.~Wang, T.~Ma, L.~Wang, J.~Chen, W.-S. Zheng, and T.~Zhang,
  ``Towards unbiased covid-19 lesion localisation and segmentation via weakly
  supervised learning,'' in \emph{IEEE International Symposium on Biomedical
  Imaging (ISBI)}.\hskip 1em plus 0.5em minus 0.4em\relax IEEE, 2021, pp.
  1966--1970.

\bibitem{wang_dofe_2020}
S.~Wang, L.~Yu, K.~Li, X.~Yang, C.-W. Fu, and P.-A. Heng, ``{DoFE}:
  {Domain}-{Oriented} {Feature} {Embedding} for {Generalizable} {Fundus}
  {Image} {Segmentation} on {Unseen} {Datasets},'' \emph{IEEE Transactions on
  Medical Imaging}, vol.~39, no.~12, pp. 4237--4248, 2020.

\bibitem{liu_feddg_2021}
Q.~Liu, C.~Chen, J.~Qin, Q.~Dou, and P.-A. Heng, ``{FedDG}: {Federated}
  {Domain} {Generalization} on {Medical} {Image} {Segmentation} via {Episodic}
  {Learning} in {Continuous} {Frequency} {Space},'' in \emph{IEEE Conference on
  Computer Vision and Pattern Recognition (CVPR)}, Jun. 2021, pp. 1013--1023.

\bibitem{yang_fda_2020}
Y.~Yang and S.~Soatto, ``{FDA}: {Fourier} {Domain} {Adaptation} for {Semantic}
  {Segmentation},'' in \emph{IEEE Conference on Computer Vision and Pattern
  Recognition (CVPR)}, Jun. 2020, pp. 4085--4095.

\bibitem{chen2020unsupervised}
C.~Chen, Q.~Dou, H.~Chen, J.~Qin, and P.~A. Heng, ``Unsupervised bidirectional
  cross-modality adaptation via deeply synergistic image and feature alignment
  for medical image segmentation,'' \emph{IEEE Transactions on Medical
  Imaging}, vol.~39, no.~7, pp. 2494--2505, 2020.

\bibitem{liu2020pdam}
D.~Liu, D.~Zhang, Y.~Song, F.~Zhang, L.~O’Donnell, H.~Huang, M.~Chen, and
  W.~Cai, ``Pdam: A panoptic-level feature alignment framework for unsupervised
  domain adaptive instance segmentation in microscopy images,'' \emph{IEEE
  Transactions on Medical Imaging}, vol.~40, no.~1, pp. 154--165, 2020.

\bibitem{liu_ms-net_2020}
Q.~Liu, Q.~Dou, L.~Yu, and P.~A. Heng, ``{MS}-{Net}: {Multi}-{Site} {Network}
  for {Improving} {Prostate} {Segmentation} {With} {Heterogeneous} {MRI}
  {Data},'' \emph{IEEE Transactions on Medical Imaging}, vol.~39, no.~9, pp.
  2713--2724, Sep. 2020.

\bibitem{shen_domain-invariant_2020}
Y.~Shen, B.~Sheng, R.~Fang, H.~Li, L.~Dai, S.~Stolte, J.~Qin, W.~Jia, and
  D.~Shen, ``Domain-invariant interpretable fundus image quality assessment,''
  \emph{Medical Image Analysis}, vol.~61, p. 101654, Apr. 2020.

\bibitem{wang2019boundary}
S.~Wang, L.~Yu, K.~Li, X.~Yang, C.-W. Fu, and P.-A. Heng, ``Boundary and
  entropy-driven adversarial learning for fundus image segmentation,'' in
  \emph{Medical Image Computing and Computer Assisted Intervention (MICCAI)},
  2019, pp. 102--110.

\bibitem{tsai_rsna_2021}
E.~B. Tsai, S.~Simpson, M.~Lungren, M.~Hershman, L.~Roshkovan, E.~Colak, B.~J.
  Erickson, G.~Shih, A.~Stein, J.~Kalpathy-Cramer, J.~Shen, M.~Hafez, S.~John,
  P.~Rajiah, B.~P. Pogatchnik, J.~Mongan, E.~Altinmakas, E.~R. Ranschaert,
  F.~C. Kitamura, L.~Topff, L.~Moy, J.~P. Kanne, and C.~C. Wu, ``The {RSNA}
  {International} {COVID}-19 {Open} {Annotated} {Radiology} {Database}
  ({RICORD}),'' \emph{Radiology}, vol. 299, no.~1, pp. E204--E213, Jan. 2021.

\bibitem{roth2021rapid}
H.~Roth, Z.~Xu, C.~T. Diez, R.~S. Jacob, J.~Zember, J.~Molto, W.~Li, S.~Xu,
  B.~Turkbey, E.~Turkbey \emph{et~al.}, ``Rapid artificial intelligence
  solutions in a pandemic-the covid-19-20 lung ct lesion segmentation
  challenge,'' \emph{Research Square}, 2021.

\bibitem{he2021autoencoder}
Y.~He, A.~Carass, L.~Zuo, B.~E. Dewey, and J.~L. Prince, ``Autoencoder based
  self-supervised test-time adaptation for medical image analysis,''
  \emph{Medical Image Analysis}, vol.~72, p. 102136, 2021.

\bibitem{karani_test-time_2021}
N.~Karani, E.~Erdil, K.~Chaitanya, and E.~Konukoglu,
  ``\BIBforeignlanguage{en}{Test-time adaptable neural networks for robust
  medical image segmentation},'' \emph{\BIBforeignlanguage{en}{Medical Image
  Analysis}}, vol.~68, p. 101907, Feb. 2021.

\bibitem{fan_adversarially_2021}
X.~Fan, Q.~Wang, J.~Ke, F.~Yang, B.~Gong, and M.~Zhou, ``Adversarially
  {Adaptive} {Normalization} for {Single} {Domain} {Generalization},'' in
  \emph{IEEE Conference on Computer Vision and Pattern Recognition (CVPR)},
  Jun. 2021, pp. 8208--8217.

\bibitem{zhou_duplex_2021}
Q.~Zhou, W.~Zhou, S.~Wang, and Y.~Xing, ``\BIBforeignlanguage{en}{Duplex
  adversarial networks for multiple-source domain adaptation},''
  \emph{\BIBforeignlanguage{en}{Knowledge-Based Systems}}, vol. 211, p. 106569,
  Jan. 2021.

\bibitem{liu_shape-aware_2020}
Q.~Liu, Q.~Dou, and P.-A. Heng, ``\BIBforeignlanguage{en}{Shape-{Aware}
  {Meta}-learning for {Generalizing} {Prostate} {MRI} {Segmentation} to
  {Unseen} {Domains}},'' in \emph{\BIBforeignlanguage{en}{Medical Image
  Computing and Computer Assisted Intervention (MICCAI)}}, 2020, pp. 475--485.

\bibitem{du_metanorm_2021}
Y.~Du, X.~Zhen, L.~Shao, and C.~G.~M. Snoek, ``{MetaNorm}: {Learning} to
  {Normalize} {Few}-{Shot} {Batches} {Across} {Domains},'' in
  \emph{International Conference on Learning Representations (ICLR)}.\hskip 1em
  plus 0.5em minus 0.4em\relax OpenReview.net, 2021.

\bibitem{liu_semi-supervised_2021}
X.~Liu, S.~Thermos, A.~O'Neil, and S.~Tsaftaris, ``Semi-supervised
  {Meta}-learning with {Disentanglement} for {Domain}-generalised {Medical}
  {Image} {Segmentation},'' in \emph{Medical Image Computing and Computer
  Assisted Intervention (MICCAI)}, Jun. 2021.

\bibitem{zhang_generalizing_2020}
L.~Zhang, X.~Wang, D.~Yang, T.~Sanford, S.~Harmon, B.~Turkbey, B.~J. Wood,
  H.~Roth, A.~Myronenko, D.~Xu, and Z.~Xu, ``Generalizing {Deep} {Learning} for
  {Medical} {Image} {Segmentation} to {Unseen} {Domains} via {Deep} {Stacked}
  {Transformation},'' \emph{IEEE Transactions on Medical Imaging}, vol.~39,
  no.~7, pp. 2531--2540, Jul. 2020.

\bibitem{li_domain_2020}
H.~Li, Y.~Wang, R.~Wan, S.~Wang, T.-Q. Li, and A.~Kot, ``Domain
  {Generalization} for {Medical} {Imaging} {Classification} with
  {Linear}-{Dependency} {Regularization},'' in \emph{Advances in Neural
  Information Processing Systems (NeurIPS)}, 2020, pp. 3115--3126.

\bibitem{gu2021domain}
R.~Gu, J.~Zhang, R.~Huang, W.~Lei, G.~Wang, and S.~Zhang, ``Domain composition
  and attention for unseen-domain generalizable medical image segmentation,''
  in \emph{Medical Image Computing and Computer Assisted Intervention
  (MICCAI)}.\hskip 1em plus 0.5em minus 0.4em\relax Springer, 2021, pp.
  241--250.

\bibitem{li_episodic_2019}
D.~Li, J.~Zhang, Y.~Yang, C.~Liu, Y.-Z. Song, and T.~M. Hospedales, ``Episodic
  {Training} for {Domain} {Generalization},'' in \emph{International Conference
  on Computer Vision (ICCV)}, Oct. 2019.

\bibitem{li_domain_2021}
C.~Li, Q.~Qi, X.~Ding, Y.~Huang, D.~Liang, and Y.~Yu, ``Domain {Generalization}
  on {Medical} {Imaging} {Classification} using {Episodic} {Training} with
  {Task} {Augmentation},'' \emph{arXiv:2106.06908 [cs]}, Jun. 2021.

\bibitem{onofrey_generalizable_2019}
J.~A. Onofrey, D.~I. Casetti-Dinescu, A.~D. Lauritzen, S.~Sarkar,
  R.~Venkataraman, R.~E. Fan, G.~A. Sonn, P.~C. Sprenkle, L.~H. Staib, and
  X.~Papademetris, ``Generalizable {Multi}-{Site} {Training} and {Testing} {Of}
  {Deep} {Neural} {Networks} {Using} {Image} {Normalization},'' in \emph{IEEE
  International Symposium on Biomedical Imaging (ISBI)}, Apr. 2019, pp.
  348--351.

\bibitem{zhao_robust_2021}
X.~Zhao, A.~Sicilia, D.~S. Minhas, E.~E. O’Connor, H.~J. Aizenstein, W.~E.
  Klunk, D.~L. Tudorascu, and S.~J. Hwang, ``Robust {White} {Matter}
  {Hyperintensity} {Segmentation} {on} {Unseen} {Domain},'' in \emph{IEEE
  International Symposium on Biomedical Imaging (ISBI)}, Apr. 2021, pp.
  1047--1051.

\bibitem{he_dynamic_2019}
J.~He, Z.~Deng, and Y.~Qiao, ``Dynamic {Multi}-{Scale} {Filters} for {Semantic}
  {Segmentation},'' in \emph{International Conference on Computer Vision
  (ICCV)}, Oct. 2019, pp. 3561--3571.

\bibitem{pang_hierarchical_2020}
Y.~Pang, L.~Zhang, X.~Zhao, and H.~Lu, ``Hierarchical {Dynamic} {Filtering}
  {Network} for {RGB}-{D} {Salient} {Object} {Detection},'' in \emph{European
  Conference on Computer Vision (ECCV)}, 2020, pp. 235--252.

\bibitem{zhou_decoupled_2021}
J.~Zhou, V.~Jampani, Z.~Pi, Q.~Liu, and M.-H. Yang, ``Decoupled {Dynamic}
  {Filter} {Networks},'' in \emph{IEEE Conference on Computer Vision and
  Pattern Recognition (CVPR)}, Jun. 2021, pp. 6647--6656.

\bibitem{han_dynamic_2021}
Y.~Han, G.~Huang, S.~Song, L.~Yang, H.~Wang, and Y.~Wang, ``Dynamic {Neural}
  {Networks}: {A} {Survey},'' \emph{arXiv:2102.04906 [cs]}, Feb. 2021.

\bibitem{klein_dynamic_2015}
B.~Klein, L.~Wolf, and Y.~Afek, ``A {Dynamic} {Convolutional} {Layer} for short
  range weather prediction,'' in \emph{IEEE Conference on Computer Vision and
  Pattern Recognition (CVPR)}, Jun. 2015, pp. 4840--4848.

\bibitem{tian_conditional_2020}
Z.~Tian, C.~Shen, and H.~Chen, ``Conditional {Convolutions} for {Instance}
  {Segmentation},'' in \emph{European Conference on Computer Vision (ECCV)},
  Aug. 2020, pp. 282--298.

\bibitem{zhang_dodnet_2021}
J.~Zhang, Y.~Xie, Y.~Xia, and C.~Shen, ``{DoDNet}: {Learning} {To} {Segment}
  {Multi}-{Organ} and {Tumors} {From} {Multiple} {Partially} {Labeled}
  {Datasets},'' in \emph{IEEE Conference on Computer Vision and Pattern
  Recognition (CVPR)}, Jun. 2021, pp. 1195--1204.

\bibitem{bloch2015nci}
N.~Bloch, A.~Madabhushi, H.~Huisman, J.~Freymann, J.~Kirby, M.~Grauer,
  A.~Enquobahrie, C.~Jaffe, L.~Clarke, and K.~Farahani, ``Nci-isbi 2013
  challenge: automated segmentation of prostate structures,'' \emph{The Cancer
  Imaging Archive}, vol. 370, p.~6, 2015.

\bibitem{lemaitre2015computer}
G.~Lema{\^\i}tre, R.~Mart{\'\i}, J.~Freixenet, J.~C. Vilanova, P.~M. Walker,
  and F.~Meriaudeau, ``Computer-aided detection and diagnosis for prostate
  cancer based on mono and multi-parametric mri: a review,'' \emph{Computers in
  Biology and Medicine}, vol.~60, pp. 8--31, 2015.

\bibitem{litjens2014evaluation}
G.~Litjens, R.~Toth, W.~van~de Ven, C.~Hoeks, S.~Kerkstra, B.~van Ginneken,
  G.~Vincent, G.~Guillard, N.~Birbeck, J.~Zhang \emph{et~al.}, ``Evaluation of
  prostate segmentation algorithms for mri: the promise12 challenge,''
  \emph{Medical Image Analysis}, vol.~18, no.~2, pp. 359--373, 2014.

\bibitem{clark2013cancer}
K.~Clark, B.~Vendt, K.~Smith, J.~Freymann, J.~Kirby, P.~Koppel, S.~Moore,
  S.~Phillips, D.~Maffitt, M.~Pringle \emph{et~al.}, ``The cancer imaging
  archive (tcia): maintaining and operating a public information repository,''
  \emph{Journal of Digital Imaging}, vol.~26, no.~6, pp. 1045--1057, 2013.

\bibitem{ricord1a}
\BIBentryALTinterwordspacing
E.~Tsai, S.~Simpson, M.~P. Lungren, M.~Hershman, L.~Roshkovan, E.~Colak, B.~J.
  Erickson, G.~Shih, A.~Stein, J.~Kalpathy-Cramer, J.~Shen, M.~A. Hafez,
  S.~John, P.~Rajiah, B.~P. Pogatchnik, J.~T. Mongan, E.~Altinmakas,
  E.~Ranschaert, F.~C. Kitamura, L.~Topff, L.~Moy, J.~P. Kanne, and C.~C. Wu,
  ``Medical imaging data resource center - rsna international covid radiology
  database release 1a - chest ct covid+ (midrc-ricord-1a),'' 2020. [Online].
  Available: \url{https://wiki.cancerimagingarchive.net/x/DoDTB}
\BIBentrySTDinterwordspacing

\bibitem{sivaswamy2015comprehensive}
J.~Sivaswamy, S.~Krishnadas, A.~Chakravarty, G.~Joshi, A.~S. Tabish
  \emph{et~al.}, ``A comprehensive retinal image dataset for the assessment of
  glaucoma from the optic nerve head analysis,'' \emph{JSM Biomedical Imaging
  Data Papers}, vol.~2, no.~1, p. 1004, 2015.

\bibitem{fumero2011rim}
F.~Fumero, S.~Alay{\'o}n, J.~L. Sanchez, J.~Sigut, and M.~Gonzalez-Hernandez,
  ``Rim-one: An open retinal image database for optic nerve evaluation,'' in
  \emph{International Symposium on Computer-based Medical Systems
  (CBMS)}.\hskip 1em plus 0.5em minus 0.4em\relax IEEE, 2011, pp. 1--6.

\bibitem{orlando2020refuge}
J.~I. Orlando, H.~Fu, J.~B. Breda, K.~van Keer, D.~R. Bathula, A.~Diaz-Pinto,
  R.~Fang, P.-A. Heng, J.~Kim, J.~Lee \emph{et~al.}, ``Refuge challenge: A
  unified framework for evaluating automated methods for glaucoma assessment
  from fundus photographs,'' \emph{Medical Image Analysis}, vol.~59, p. 101570,
  2020.

\bibitem{chen2018encoder}
L.-C. Chen, Y.~Zhu, G.~Papandreou, F.~Schroff, and H.~Adam, ``Encoder-decoder
  with atrous separable convolution for semantic image segmentation,'' in
  \emph{European Conference on Computer Vision (ECCV)}, 2018, pp. 801--818.

\bibitem{chen_cooperative_2021}
C.~Chen, K.~Hammernik, C.~Ouyang, C.~Qin, W.~Bai, and D.~Rueckert,
  ``Cooperative training and latent space data augmentation for robust medical
  image segmentation,'' in \emph{Medical Image Computing and Computer Assisted
  Intervention (MICCAI)}.\hskip 1em plus 0.5em minus 0.4em\relax Springer,
  2021, pp. 149--159.

\bibitem{ouyang_causality-inspired_2021}
C.~Ouyang, C.~Chen, S.~Li, Z.~Li, C.~Qin, W.~Bai, and D.~Rueckert,
  ``Causality-inspired single-source domain generalization for medical image
  segmentation,'' \emph{arXiv:2111.12525 [cs]}, Dec. 2021.

\bibitem{hu_domain_2022}
S.~Hu, Z.~Liao, and Y.~Xia, ``Domain specific convolution and high frequency
  reconstruction based unsupervised domain adaptation for medical image
  segmentation,'' in \emph{Medical Image Computing and Computer Assisted
  Intervention (MICCAI)}, 2022, pp. 650--659.

\bibitem{decenciere2014feedback}
E.~Decenci{\`e}re, X.~Zhang, G.~Cazuguel, B.~Lay, B.~Cochener, C.~Trone,
  P.~Gain, R.~Ordonez, P.~Massin, A.~Erginay \emph{et~al.}, ``Feedback on a
  publicly distributed image database: the messidor database,'' \emph{Image
  Analysis \& Stereology}, vol.~33, no.~3, pp. 231--234, 2014.

\bibitem{almazroa2018retinal}
A.~Almazroa, S.~Alodhayb, E.~Osman, E.~Ramadan, M.~Hummadi, M.~Dlaim,
  M.~Alkatee, K.~Raahemifar, and V.~Lakshminarayanan, ``Retinal fundus images
  for glaucoma analysis: the riga dataset,'' in \emph{Medical Imaging 2018:
  Imaging Informatics for Healthcare, Research, and Applications}, vol.
  10579.\hskip 1em plus 0.5em minus 0.4em\relax International Society for
  Optics and Photonics, 2018, p. 105790B.

\end{thebibliography}

\clearpage

\appendix
\setcounter{table}{0}
\renewcommand{\thetable}{A\arabic{table}}

\subsection*{Analysis of Domain Code}
To analyze whether the domain code match the similarities between the test image and $K$ source domains, the images from Domain 4 were chosen as the target domain for a case study. 
We used the DCAC model trained on the data from $K$ source domains (except for Domain 4) to produce the domain code for target domain images. 
Since the domain code gives the ranking of similarities between a test image and source domains, Table~\ref{tab:a4} shows the percentage of the ranking of all test images over five source domains. 
For instance, 95.43\% test images are most similar to Domain 1, and 69.00\% test images are second-most similar to Domain 2. 
The observation that most test images are classified as Domain 1 or Domain 2 confirms that the domain code does match the similarities between the test image and source domains.

\begin{table}[]
\centering
\caption{Percentage of similarity ranking of all test images over five source domains according to domain code (using Domain 4 as target domain).}
\label{tab:a4}
\renewcommand{\arraystretch}{1.2} 
\setlength\tabcolsep{4pt}
\begin{tabular}{l|ccccc}
\hline \hline
 & Rank-1 & Rank-2 & Rank-3 & Rank-4 & Rank-5 \\
\hline
Domain 1 & 4.57 & 69.00 & 12.15 & 14.29 & 0.00 \\
Domain 2 & 95.43 & 3.71 & 0.86 & 0.00 & 0.00 \\
Domain 3 & 0.00 & 5.00 & 81.85 & 2.29 & 10.86 \\
Domain 5 & 0.00 & 7.14 & 0.86 & 69.42 & 22.58 \\
Domain 6 & 0.00 & 15.14 & 4.29 & 14.00 & 66.57 \\
\hline \hline
\end{tabular}
\end{table}

\subsection*{Experiments on Other Multi-domain Dataset}
We also evaluated our DCAC model on another multi-domain joint OC/OD segmentation dataset called RIGA+~\cite{hu_domain_2022,decenciere2014feedback,almazroa2018retinal}, which has larger domain gaps. This dataset contains 195 labeled data from BinRushed, 95 labeled data from Magrabia, and 454 labeled data from the MESSIDOR database (including data collected from 3 medical centers, \textit{i.e.}, BASE1, BASE2, and BASE3). We used the data from BinRushed and Magrabia as source domain data, and the data from BASE1, BASE2, and BASE3 as target domain data, respectively. We compared DCAC with the `Intra-domain' setting and the `DeepAll' baseline. Table~\ref{tab:a5} gives the performance ($DSC_{OC}$\%,$DSC_{OD}$\%) of these three methods. It shows that the domain gap indicated by the performance difference between `Intra-domain' and `DeepAll' is severe, especially on BASE2, \textit{i.e.}, a decrease of Dice by 9.61\% in OC segmentation. Even though, our DCAC can improve the OC segmentation Dice by 3.49\% (from 79.22\% to 82.71\%) on BASE2 without accessing any target domain data. 
It confirms the effectiveness of the proposed DCAC. 

\begin{table}[]
\centering
\caption{Performance of `Intra-domain', `DeepAll', and DCAC on RIGA+ dataset.}
\label{tab:a5}
\renewcommand{\arraystretch}{1.2} 
\setlength\tabcolsep{4pt}
\begin{tabular}{l|ccc}
\hline \hline
Models & BASE1 & BASE2 & BASE3 \\
\hline
Intra-domain & (87.27, 95.87) & (88.83, 95.92) & (88.42, 95.73) \\
DeepAll & (84.01, 94.30) & (79.22, 94.58) & (81.40, 93.82) \\
\hline
DCAC & (85.00, 95.86) & (82.71, 95.20) & (83.09, 95.10) \\
\hline \hline
\end{tabular}
\end{table}

\subsection*{Analysis of Residual Learning in the DAC Head}
For Formula (5) in Sec. III-C-3, using `$-$' and using `$+$' are equivalent, since we did not constrain that $\omega_d$ must be positive when generating it using the domain-aware controller. 
We compared the performance of DCAC$+$ (\textit{i.e.}, using `$+$' for residual learning) and DCAC (\textit{i.e.}, using `$-$' for residual learning) in Table~\ref{tab:a1}. It shows that using `$+$' for residual learning can still work well.

\begin{table*}[]
\centering
\caption{Performance of using `$+$' (DCAC$+$) and `$-$' (DCAC) for residual learning.}
\label{tab:a1}
\renewcommand{\arraystretch}{1.2} 
\setlength\tabcolsep{4pt}
\begin{tabular}{l|cccccccccccc|cc}
\hline \hline
\multirow{2}{*}{Models} & \multicolumn{2}{c}{Domain 1} & \multicolumn{2}{c}{Domain 2} & \multicolumn{2}{c}{Domain 3} & \multicolumn{2}{c}{Domain 4} & \multicolumn{2}{c}{Domain 5} & \multicolumn{2}{c}{Domain 6} & \multicolumn{2}{|c}{Average} \\ 
\cline{2-15}
          & DSC$\uparrow$   & ASD$\downarrow$   & DSC$\uparrow$   & ASD$\downarrow$   & DSC$\uparrow$   & ASD$\downarrow$   & DSC$\uparrow$   & ASD$\downarrow$   & DSC$\uparrow$   & ASD$\downarrow$   & DSC$\uparrow$   & ASD$\downarrow$   & DSC$\uparrow$   & ASD$\downarrow$   \\ \hline
DCAC$+$ & 91.61 & 1.09 & 90.52 & 0.88 & 86.52 & 1.60 & 89.20 & 1.45 & 82.74 & 2.47 & 90.24 & 0.96 & 88.47 & 1.41 \\
DCAC & 91.76 & 0.98 & 90.51 & 0.89 & 86.30 & 1.77 & 89.13 & 1.53 & 83.39 & 2.46 & 90.56 & 0.85 & 88.61 & 1.41 \\ 
\hline \hline
\end{tabular}
\end{table*}

\subsection*{Analysis of Multi-scale Supervision}
As for the multi-scale supervision, we use the DAC head and CAC head at each scale in the decoder. We conducted an ablation study on the number of scales. Let $N-1$ is the scales of the decoder. Since selecting the number of scales requires to conduct $2^{N-1}$ experiments, which is computationally intractable, we only compared the performance of DCAC$^n$ (\textit{i.e.}, using DAC and CAC at the last $n$ decoder blocks) with the performance of DCAC. The results were shown in Table~\ref{tab:a2}. It reveals that our DCAC model that uses DAC and CAC at all scales in the decoder outperforms all competing ones. It confirms the rationale and effectiveness of our design of the deep supervision strategy.

\begin{table*}[]
\centering
\caption{Performance of DCAC and DCAC$^n$.}
\label{tab:a2}
\renewcommand{\arraystretch}{1.2} 
\setlength\tabcolsep{4pt}
\begin{tabular}{l|cccccccccccc|cc}
\hline \hline
\multirow{2}{*}{Models} & \multicolumn{2}{c}{Domain 1} & \multicolumn{2}{c}{Domain 2} & \multicolumn{2}{c}{Domain 3} & \multicolumn{2}{c}{Domain 4} & \multicolumn{2}{c}{Domain 5} & \multicolumn{2}{c}{Domain 6} & \multicolumn{2}{|c}{Average} \\ 
\cline{2-15}
          & DSC$\uparrow$   & ASD$\downarrow$   & DSC$\uparrow$   & ASD$\downarrow$   & DSC$\uparrow$   & ASD$\downarrow$   & DSC$\uparrow$   & ASD$\downarrow$   & DSC$\uparrow$   & ASD$\downarrow$   & DSC$\uparrow$   & ASD$\downarrow$   & DSC$\uparrow$   & ASD$\downarrow$   \\ \hline
DCAC$^1$ & 91.02 & 1.28 & 89.93 & 0.95 & 84.55 & 2.01 & 88.52 & 2.02 & 72.58 & 4.27 & 90.41 & 0.92 & 86.17 & 1.91 \\
DCAC$^2$ & 91.42 & 1.31 & 90.01 & 0.93 & 85.28 & 1.59 & 89.15 & 1.82 & 78.38 & 5.78 & 90.95 & 0.84 & 87.53 & 2.05 \\
DCAC$^3$ & 90.83 & 1.55 & 90.46 & 0.90 & 85.79 & 1.78 & 89.16 & 1.71 & 79.97 & 4.64 & 90.34 & 0.97 & 87.76 & 1.92 \\
DCAC$^4$ & 91.44 & 1.15 & 90.35 & 0.90 & 86.28 & 1.64 & 89.40 & 1.70 & 80.19 & 3.20 & 90.54 & 0.87 & 88.03 & 1.58 \\
DCAC$^5$ & 91.10 & 1.22 & 90.48 & 0.89 & 85.77 & 1.76 & 89.63 & 1.56 & 82.11 & 2.30 & 90.35 & 0.94 & 88.24 & 1.45 \\
DCAC & 91.76 & 0.98 & 90.51 & 0.89 & 86.30 & 1.77 & 89.13 & 1.53 & 83.39 & 2.46 & 90.56 & 0.85 & 88.61 & 1.41 \\ 
\hline \hline
\end{tabular}
\end{table*}

\subsection*{Analysis of Domain-discriminatory Power of Multi-scale Features}
We also compared the performance of DCAC when using the global image features from the bottleneck (\textit{i.e.}, DCAC) and using the features from multiple encoder blocks (denoted by DCAC-M) as the input of the CAC module. Table~\ref{tab:a3} shows the experimental results. It reveals that using global image features from the bottleneck is slightly better than using the features from multiple encoder blocks. We speculate that it should be attributed to the fact that there are redundant features in the aggregated multi-scale feature, which might be not content-specific. 

\begin{table*}[]
\centering
\caption{Performance of DCAC and DCAC-M.}
\label{tab:a3}
\renewcommand{\arraystretch}{1.2} 
\setlength\tabcolsep{4pt}
\begin{tabular}{l|cccccccccccc|cc}
\hline \hline
\multirow{2}{*}{Models} & \multicolumn{2}{c}{Domain 1} & \multicolumn{2}{c}{Domain 2} & \multicolumn{2}{c}{Domain 3} & \multicolumn{2}{c}{Domain 4} & \multicolumn{2}{c}{Domain 5} & \multicolumn{2}{c}{Domain 6} & \multicolumn{2}{|c}{Average} \\ 
\cline{2-15}
          & DSC$\uparrow$   & ASD$\downarrow$   & DSC$\uparrow$   & ASD$\downarrow$   & DSC$\uparrow$   & ASD$\downarrow$   & DSC$\uparrow$   & ASD$\downarrow$   & DSC$\uparrow$   & ASD$\downarrow$   & DSC$\uparrow$   & ASD$\downarrow$   & DSC$\uparrow$   & ASD$\downarrow$   \\ \hline
DCAC-M & 91.64 & 1.05 & 90.52 & 0.88 & 85.82 & 1.67 & 89.56 & 1.19 & 80.02 & 2.99 & 90.71 & 0.90 & 88.04 & 1.45 \\
DCAC & 91.76 & 0.98 & 90.51 & 0.89 & 86.30 & 1.77 & 89.13 & 1.53 & 83.39 & 2.46 & 90.56 & 0.85 & 88.61 & 1.41 \\ 
\hline \hline
\end{tabular}
\end{table*}

\end{document}